\documentclass[10pt]{IEEEtran}

\usepackage{amsfonts}
\usepackage{bbm}
\usepackage{bm}
\usepackage{slashbox}
\usepackage{booktabs}
\usepackage{threeparttable}
\usepackage{amsfonts}
\usepackage{amssymb}
\usepackage{cite}
\usepackage{graphicx}
\usepackage{subfigure}
\usepackage{url}
\usepackage{stfloats}
\usepackage{amsmath}
\usepackage{mathtools}
\usepackage{array}

\usepackage{mathrsfs}

\usepackage{multirow}

\usepackage{color}

\usepackage{xcolor}
\usepackage{bbm}

\usepackage[justification=centering]{caption}

\ifCLASSINFOpdf
\else
\fi
\begin{document}

%\title{ Finite Frequency Memory  Output  Feedback Controller Design for T-S Fuzzy Affine Systems
%\thanks {This work was supported in part by the Research Grants Council of the Hong Kong Special Administrative Region of China (No. CityU/11203714), the National Natural Science Foundation of China (61522306), and the State Key Laboratory of Synthetical Automation for Process Industries (Northeastern University, Shenyang, China).}}

%\author{Meng Wang, Gang Feng, \IEEEmembership{Fellow,~IEEE,} and Jianbin Qiu, \IEEEmembership{Senior Member,~IEEE}
%\thanks{M. Wang and G. Feng are with the Department of Mechanical and Biomedical
%Engineering, City University of Hong Kong, Kowloon, Hong Kong.
%Email: mwang48-c@my.cityu.edu.hk (M. Wang), megfeng@cityu.edu.hk (G. Feng).}
%\thanks{J. Qiu is with the Research Institute of Intelligent Control and Systems, Harbin Institute of Technology, Harbin 150080, P. R. China. Email: jbqiu@hit.edu.cn (J. Qiu).}
%}

\title{ Memory-Based Control with Event-Triggered Protocol for interval type-2 fuzzy network system under fading channel
% Memory output-feedback controller design with Memory-based Dynamic Event-Triggered Protocol for interval type-2 fuzzy network system under fading channel
% \thanks {This work was supported in part by the Research Grants Council of the Hong Kong Special Administrative Region of China (No. CityU/11204315),
%   the National Natural Science Foundation of China (61522306), and the State Key Laboratory of Synthetical Automation for Process Industries (Northeastern University, Shenyang, China).}
}

\author{Sen Kong, Meng Wang
% \thanks{M. Wang and G. Feng are with the Department of Mechanical and Biomedical
% Engineering, City University of Hong Kong, Kowloon, Hong Kong.
% Email: mwang48-c@my.cityu.edu.hk (M. Wang), megfeng@cityu.edu.hk (G. Feng).}
% \thanks{J. Qiu is with the Research Institute of Intelligent Control and Systems, Harbin Institute of Technology, Harbin 150080, P. R. China. Email: jbqiu@hit.edu.cn (J. Qiu).}
}

\maketitle

\begin{abstract}
To address the challenges in networked environments and control problems associated with complex nonlinear uncertain systems, this paper investigates the design of a membership-function-dependent (MFD) memory output-feedback (MOF) controller for interval type-2 (IT2) fuzzy systems under fading channels, leveraging a memory dynamic event-triggering mechanism (MDETM). To conserve communication resources, MDETM reduces the frequency of data transmission. For mitigating design conservatism, a MOF controller is employed. A stochastic process models the fading channel, accounting for phenomena such as reflection, refraction, and diffraction that occur during data packet transmission through networks. An actuator failure model addresses potential faults and inaccuracies in practical applications. Considering the impacts of channel fading and actuator failures, the non-parallel distributed compensation (non-PDC) strategy enhances the robustness and anti-interference capability of the MDETM MOF controller. By fully exploiting membership function information, novel MFD control design results ensure mean-square exponential stability and $\mathscr H_{\infty}$ performance $\gamma$ for the closed-loop system. Simulation studies validate the effectiveness of the proposed approach.
\end{abstract}


%%%%%%%%%%%%%%%%%%%%%%%%%%%%%%%%%%%%%%%%%%%%%%%%%%%%%%%%%%%%%%%%%%%%%%%%%%%%%%%%%%%%%%%%%
\section{Introduction}
It has been well recognized that almost all industrial processes and physical plants are nonlinear in practice, and the control strategy based on fuzzy model is one of the most effective control methods for complex nonlinear systems [1], [2]. In particular, the famous method based on the Takagi-Sugeno (T-S) fuzzy model has been widely studied due to its powerful nonlinear function approximation ability [3], [7].

However, in [8], it was found that a type-1 T-S fuzzy model with crisp membership functions fails to capture the parameter uncertainty in nonlinear plants, and IT2 T-S fuzzy models can represent general uncertain nonlinear systems because of the not crisp membership functions. In recent years, the interest in stability and synthesis techniques of IT2 T-S fuzzy models has increased rapidly[9]–[11]. For instance, the authors of [9] obtained the stability analysis LMIs conditions for the IT2 T-S fuzzy systems, the authors of [10] deals with the problems of stochastic asymptotic stability analysis in the Ito stochastic-delayed IT2 fuzzy systems by using line integral type Lyapunov-Krasovskii functional, and the authors of [11] achieved the membership-function-dependent non-parallel distributed compensation controller design method.

Generally speaking, it can be understood that the T-S fuzzy control system divides the closed-loop control system into two parts, linear (linear control subsystems) and nonlinear (membership functions). That makes stability analysis become easy by mainly studying linear parts through all permutations of membership functions. However, the information about the membership functions is not considered in the stability analysis, which explains the source of conservation. Sometimes, the analysis strategy of the original nonlinear model may provide better solutions than that of the equivalent T-S fuzzy one [12]. This observation provides a method by introducing information of membership functions into stability conditions to relaxing stability conditions, and motivating the MFD stability analysis. As mentioned above, the membership function information plays an important role in the relaxation of stability analysis and controller synthesis results, including the literature [12]-[14]. The author in [13] summarized that the MFD stability analysis approaches can introduce the global boundary information, the local boundary information, or the approximate membership function into the stability analysis. In [14], the new relaxation stability condition was derived by using the information of the uncertain membership function, with the known minimum and maximum rank of membership. By incorporateing the membership function shape information in the form of polynomial constraints, a series of relaxed linear matrix inequalities (LMIs) conditions were given in [12].

In the direction of reducing design conservation, a recent approach called memory control has attracted much attention of scholars, with some types of memory control methods have developed in [15] - [18]. Compared to the traditional memoryless control methods, the memory control methods have obvious advantages, including lower conservation and better control performance. Mention that, the authors of [15], [16] and [17] utilized the historical states of the system to propose periodic time-varying memory state feedback and static output feedback controller design methods, which regard memorless state feedback controller and static output feedback controller as special cases of memory controller, respectively. In [18], the authors developed the so-called exact-memory control approach in delay systems.

On the other hand, in the past two decades, network control system (NCS) has been a hot research field, a lot of outstanding research works have been published, and the relevant research results of NCSs have been widely applied to practical systems, Including unmanned robot cluster systems, mobile communication, smart grid, and industrial automation. In practical NCSs, there is a critical problem to solve is reducing the burden of communication. In [19]–[23], to reduce the redundant signals transmitted in the limited network bandwidth, event-triggering mechanisms (ETMs) were proposed that can minimize the transmission of redundant signals in the control loop while guaranteeing the system performance. In fact, this mechanism is very important, especially when the batteries have limited energy supply, and too much redundant transmission will deplete the batteries. Compared to the traditional time-triggered mechanism, the ETMs can not only further reduce the packet transmission, and reduce the controller's computational burden but also affect the system performance to a minimum extent. A large number of results related to ETMs can be found in the literature, for example, [19]–[23]. The ETM in the literature [19] - [21] can be considered as a static event-triggering mechanism (STEM) because the trigger threshold is a fixed constant. The ETM in the literature [22] - [23] can be categorized as a dynamic event-triggering mechanism (DETM) because it introduces a state-dependent dynamic auxiliary variable as a trigger threshold, and since the threshold can be adaptively adjusted, the total number of triggered time could be further reduced.

It is worth noting that the ETMs in the above literature all use the latest trigger moment system states and the current ones to determine whether the trigger condition is satisfied. Nevertheless, applying the system states at specified instants may cause some unexpected triggering events. It must be pointed out that the results and applications of memory event-triggering mechanisms (METM) using historical system states are rarely discussed. As paper [24,25] are shown that considering the historical triggered data may enable the trigger mechanism to transmit more data near the concave or convex point of the response curve to shorten the system fluctuating time. In addition, using the historical state information of the system is beneficial to determine whether some special moments should be triggered.

It is also worth noting that when data packets are transmitted through the network, various phenomena would inevitably occur, such as reflection, refraction, diffraction, etc., which will lead to the channel fading. Researchers in related fields usually model channel fading as a stochastic process that affects the phase and amplitude of the transmitted data. So far, some relevant results have been reported, for example, [26] solves the control problems under channel fading, and [27] solves the filtering problems. However, to the authors' knowledge, there are only few results in the existing open literature about using METMs to design a MOF controller in networks with channel fading. All of the above statements motivate our current work.

In this paper, a MOF control problem based on a MDETM is studied for a class of IT2 fuzzy systems with fading channels. The controller effectively utilizes past and current data information in order to achieve lower conservation and better control performance. The ETM considers historical system state, while introducing a dynamic auxiliary variable, hoping to further save the limited communication bandwidth. Meanwhile, the possible channel fading phenomenon is considered in the controller design process. Moreover, to reduce the design conservationby, a novel MFD controller design method was proposed in the article. The main technical contributions can be highlighted as follows.

%A MDETM MOF control method is given to co-design the controller gain and the event-trigger parameters, while reducing both the controller conservation and the number of event triggered;
%Considering the effects of channel fading and actuator failure, adopting the nonparallel compensation strategy, and utilizing a novel MFD method, the controller design conditions are synthesized with lower resistance and stronger anti-interference;
%With the proposed controllers, the mean-square exponential stability and $H_{\infty}$ performance have been guaranteed.
%那就开始总结

%or
%The main technical contributions of the current paper are summarized as follows.
(1) A MDETM MOF control method is given to co-design the controller gain and the event-trigger parameters, while reducing both the controller conservation and the number of event triggered;

(2) Considering the effects of channel fading and actuator failure, adopting the nonparallel compensation strategy, and utilizing a novel MFD method, the controller design conditions are synthesized with lower resistance and stronger anti-interference;

(3) With the proposed controllers, the mean-square exponential stability and $\mathscr H_{\infty}$ performance have been guaranteed.

%具体安排
%The rest of the paper is organized as follows. Section 2 formulates the METS and dynamic model of the NCSs. The main results are given in Section 3, two theorems and one corollary are given for the stability and controller design of the NCSs. A simulation example is given in Section 4 and this paper is concluded in Section 5.

\textbf{Notations.}
$\Re^{n}$ denotes $n$-dimensional Euclidean space.
For a matrix $Q$, $Q^{T}$ denotes its transpose.
$Sym\{Q\}$ denotes $Q + Q^{T}$.
$Q > 0$ $\left( Q \geq 0 \right)$ or $Q < 0$ $\left( Q \leq 0 \right)$ denote $Q$ is a positive (semi-positive) definite or negative (semi-negative) definite matrix, respectively.
%For matrices $Q_{1}$, $Q_{2}$, $\cdots$, $Q_{n}$,
$diag\left\{Q_{1},\ Q_{2},\ \cdots,\ Q_{n}\right\}$ denotes the block-diagonal matrix with the block-diagonal elements $Q_{1}$, $Q_{2}$, $\cdots$, $Q_{n}$.
In a symmetric matrix, $\star$ denotes the terms that can be induced by symmetry.
%Denote $\mathscr{T}=\{-d,-d+1,\cdots,-1,0\}$, and $d$ is a positive integer.
$\|Q\|$ denotes the spectral norm of $Q$. %the matrix
$\mathbb{E}\{Q\}$ denotes the mathematical expectation of $Q$.
$\lambda(Q)$ denotes the spectral radius of the matrix $Q$.
%$\mathscr{T}$

%%%%%%%%%%%%%%%%%%%%%%%%%%%%%%%%%%%%%%%%%%%%%%%%%%%%%%%%%%%%%%%%%%%%%%%%%%%%%%%%%%%%%%%%%

%%%%%%%%%%%%%%%%%%%%%%%%%%%%%%%%%%%%%%%%%%%%%%%%%%%%%%%%%%%%%%%%%%%%%%%%%%%%%%%%%%%%%%%%%%%%%%%%%%%%%%%%%%%%%%%%%%%%%%%%%%%%%%%%%%%%%%%%%

%\maketitle \thispagestyle{plain}
%\pagestyle{empty}
%%%%%%%%%%%%%%%%%%%%%%%%%%%%%%%%%%%%%%%%%%%%%%%%%%%%%%%%%%%%%%%%%%%%%%%%%%%%%%%%%%%%%%%%%
\section{Model Description and Problem Formulation}

\subsection{Discrete-Time IT2 T-S Fuzzy Systems:}

Consider a type of discrete-time IT2 T-S fuzzy systems, which is described by the following IF-THEN rules
\begin{align}
&\textbf{Plant Rule } \mathscr{P}^{i}\textbf{: }\textbf{IF } \chi_{1}(x(t)) \textbf{ is } {M}_{1}^{i}\textbf{ and } \chi_{2}(x(t)) \textbf{ is }{M}_{2}^{i} \nonumber\\
& \textbf{ and } \cdots \textbf{ and } \chi_{\omega}(x(t))  \textbf{ is } {M}_{\omega}^{i}, \textbf{ THEN } \nonumber\\
&\begin{cases}
x(t+1)=A_{i} x(t) +{B_u}_{i}u(t) + {B_d}_{i} d(t) \\
y(t)={C_y}_{i} x(t) \\
z(t)={C_z}_{i} x(t),  \ i\in\mathscr I=\left\{1,2,\cdot\cdot\cdot,\mathfrak{p}\right\}
\label{q1}
\end{cases}
\end{align}
where
$\mathscr{P}^{i}$ symbolizes the $i$-th fuzzy inference rule, and $\mathfrak{p}$ denotes the number
of IF-THEN rules;
$\chi(x(t))=\left[\chi_{1}(x(t)), \chi_{2}(x(t)), \ldots, \chi_{\omega}(x(t))\right]$ are measurable premise variables of the system;
$M_{\psi}^{i}(\psi=1,2, \ldots, \omega)$ are type-2 fuzzy sets;
$x(t) \in \Re^{n_{x}}$, $u(t) \in \Re^{n_{u}}$, $y(t) \in \Re^{n_{y}}$, $z(t) \in \Re^{n_{z}}$, and $d(t) \in \Re^{n_{d}}$ represent the system state, the control input, the measured output, the regulated output, and the external disturbance, respectively;
$A_i$, ${B_u}_{i}$, ${B_d}_{i}$, ${C_y}_{i}$, and ${C_z}_{i}$ are known real constant matrices, which are used to represent the $i$-th local model;
$p$ is the number of inference rules.

The weight of the $i$-th rule is belonging to the interval sets:
\begin{align}
\mathscr{M}_{i}=\left[
\begin{array}{c c}
\underline{m}_{i}(x(t)) & \bar{m}_{i}(x(t))
\end{array}\right], \ i \in \mathscr{I}
\end{align}
with
\begin{equation}
\begin{gathered}
\underline{m}_{i}(x(t))=\prod_{\psi=1}^{\omega} \underline{\varsigma}_{M_{\psi}^{i}(\chi(x(t)))} \geq 0, \\
\bar{m}_{i}(x(t))=\prod_{\psi=1}^{\omega} \bar{\varsigma}_{M_{\psi}^{i}(\chi(x(t)))} \geq 0, \\
\bar{\varsigma}_{M_{\psi}^{i}(\chi(x(t)))} \geq \underline{\varsigma}_{M_{\psi}^{i}(\chi(x(t)))} \geq 0, \\
\bar{m}_{i}(x(t)) \geq \underline{m}_{i}(x(t))  \geq 0
\end{gathered}
\end{equation}
where $\underline{m}_{i}(x(t))$ and $\bar{m}_{i}(x(t))$ are the lower and upper membership functions, respectively;
$\underline{\varsigma}_{M_{\psi}^{i}}\left(\chi_{\psi}(x(t))\right)$ and $\bar{\varsigma}_{M_{\psi}^{i}(\chi(x(t)))}$ being the lower and upper grades of the membership of $\chi_{\psi}(x(t))$
in $M_{\psi}^{i}$, respectively.

Let $m_{i}(x(t))$ represent the normalized membership function satisfying
\begin{equation}
\begin{aligned}
m_{i}&(x(t))\\
=&\frac{\underline{\varphi}_{i}(x(t)) \underline{m}_{i}(x(t))+\bar{\varphi}_{i}(x(t)) \bar{m}_{i}(x(t))}{\sum_{k \in \mathscr{I}}\left(\underline{\varphi}_{k}(x(t)) \underline{m}_{k}(x(t))+\bar{\varphi}_{k}(x(t)) \bar{m}_{k}(x(t))\right)},\\
& \ \ \ \ \ \ \ \ \ \ m_{i}(x(t)) \geq 0,\ \sum_{i \in \mathscr{I}} m_{i}(x(t))=1
\end{aligned}
\end{equation}
with $\underline{\varphi}_{i}(x(t))$, $\bar{\varphi}_{i}(x(t))$ are satisfying
\begin{equation}
\begin{aligned}
0 \leq \underline{\varphi}_{i}(x(t)),  \bar{\varphi}_{i}(x(t)) \leq 1, \\
\underline{\varphi}_{i}(x(t))+\bar{\varphi}_{i}(x(t))=1.
\end{aligned}
\end{equation}

It is worth mentioning that the nonlinear functions $\underline{\varphi}_{i}(x(t))$ and $\bar{\varphi}_{i}(x(t))$ describe the parameter uncertainties, and they are unessential to be known.

%Utilizing  a center-average defuzzifier, product inference, and singleton fuzzifier, yields the following global T-S fuzzy model
The global fuzzy model can be inferred as follows:
\begin{equation}\label{Pre1}
\left\{\begin{aligned}
x(t+1)=& \sum_{i \in \mathscr{I}} m_{i}(x(t)) \left( A_{i} x(t)+ {B_u}_{i} u(t) \right. \\
& \left.+ {B_d}_{i} d(t) \right) \\
y(t)=& \sum_{i \in \mathscr{I}} m_{i}(x(t)) {C_y}_{i} x(t) \\
z(t)=&\sum_{i \in \mathscr{I}} m_{i}(x(t)) {C_z}_{i} x(t)
\end{aligned}\right. .
\end{equation}

\subsection{MDETM Subject to Fading Channel:}

%\begin{figure}[!ht]
%\centering
%\includegraphics[width=7.5cm]{3.png}\\
%\includegraphics[width=3.5in]{NCSsStructure3.eps}\\
%\caption{\footnotesize Network structure under MDETM, fading channel, MOF controller, and actuator failure with $\tilde{y}\left( t_k , \kappa \right)=[y\left(t_k\right)  \cdots y\left(t_k - h + 1 \right)  \cdots  y\left(t_k - \kappa + 1 \right)]$, $\tilde{y}_{\xi}\left( t_k , \kappa \right)=[ y_{\xi}\left(t_k\right)  \cdots  y_{\xi}\left(t_k - h + 1 \right)  \cdots  y_{\xi}\left(t_k - \kappa + 1 \right)]$}
%\end{figure}

\begin{figure*}[!ht]
    \centering
    \includegraphics[width=\textwidth]{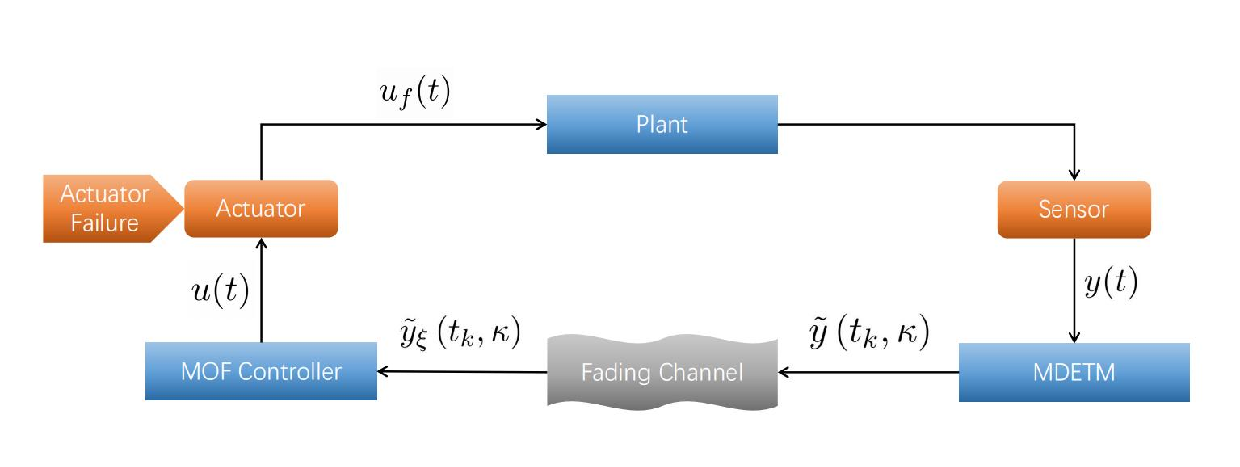}
    \caption{ Network structure under MDETM, fading channel, MOF controller, and actuator failure with $\tilde{y}\left( t_k , \kappa \right)=[y\left(t_k\right)  \cdots y\left(t_k - h + 1 \right)  \cdots  y\left(t_k - \kappa + 1 \right)]$, $\tilde{y}_{\xi}\left( t_k , \kappa \right)=[ y_{\xi}\left(t_k\right)  \cdots  y_{\xi}\left(t_k - h + 1 \right)  \cdots  y_{\xi}\left(t_k - \kappa + 1 \right)]$}
    \label{fig:network_structure} % 添加标签以便引用
\end{figure*}

To relieve the communication burden, a pretty straightforward idea is reducing the number of triggering event of data releasing.
Therefore, a MDETM, as depicted in Fig. 1, is used to estimate whether the data of output $y(t)$ should be released.

Before introducing the MDETM, the discrete DETM proposed in [ZhangZhina] is reviewed in advance.
Determine the triggered time sequence as $\{t_{k},k=0, 1, 2, \cdots ,\}$.
Denote the difference between the current output $y(t)$ and the latest released output $y(t)$ as $\varepsilon(t)=y(t)-y(t_{k})$.
The next triggered time can be decided by the following DETM condition:
\begin{equation}\begin{aligned}\label{DETM1}
t_{k+1}=&min\left\{ t|t>t_{k}, \frac{1}{\nu} \varpi(t) + \rho{y}^{T}(t) y(t) \right.\\&\left. - \varepsilon^{T}(t)\varepsilon(t) \leq 0 \right\}
\end{aligned}\end{equation}
with the auxiliary variable $\varpi(t)$ computing by
\begin{equation}\label{DETM2}
\varpi(t+1)=\mu\varpi(t) + \rho{y}^{T}(t) y(t) -  \varepsilon^{T}(t)\varepsilon(t)
\end{equation}
where $\rho$, $\nu$, and $\mu$ are given threshold parameters satisfying $0<\rho<1$, $\nu>0$, and $0<\mu<1$; the initial value of $\varpi(t)$ should satisfy $\varpi(0)\geq0$.

\label{lemma 1}
\textbf{Lemma 1. \cite{XGe2019}}  %\textup{[ZhangZhina13]}
If the scalars $\nu$ and $\mu$ in (\ref{DETM1}) and (\ref{DETM2}) satisfy
\begin{equation}
\nu\mu\geq1,
\end{equation}
it can be guaranteed that $\varpi(t)\geq0, \forall t$.

Different from the above DETM, the information of some recent system states are utilized in the proposed MDETM, which could be showen as following:
\begin{equation}\begin{aligned}\label{Pre2}
t_{k+1}=&min\left\{ t|t>t_{k}, \frac{1}{\nu} \varpi(t) + \rho{y}^{T}(t)\Omega y(t) \right.\\&\left. -  \hat{\varepsilon}^{T}(t)\Omega\hat{\varepsilon}(t) \leq 0 \right\},
\end{aligned}\end{equation}
\begin{equation}\label{Pre3}
\varpi(t+1)=\mu\varpi(t) + \rho{y}^{T}(t)\Omega y(t) - \hat{\varepsilon}^{T}(t)\Omega\hat{\varepsilon}(t)
\end{equation}
with
\begin{equation}
\left\{\begin{aligned}
& \hat{\varepsilon}(t) = \tilde{\varrho} \tilde{\varepsilon}(t) ,\\
& \tilde{\varrho}      =
    \left[\begin{array}{ccccc}
        \varrho_{\kappa}I_{n_x} & \cdots & \varrho_{h}I_{n_x} & \cdots & \varrho_{1}I_{n_x}
    \end{array}\right], \\
& \tilde{\varepsilon}(t) =
    \left[
        \varepsilon^{T}(t-\kappa+1) \ \cdots \ \varepsilon^{T}(t-h+1) \right.\\& \ \  \ \ \ \ \ \ \ \left. \cdots \ \varepsilon^{T}(t)
    \right]^{T}
%    \left[\begin{array}{ccccc}
%        \varepsilon^{T}(t-\kappa+1) & \cdots & \varepsilon^{T}(t-h+1) & \cdots & \varepsilon^{T}(t)
%    \end{array}\right]^{T}
\end{aligned}\right.
\end{equation}
where $h$ is the serial number of historical data;
$\kappa$ is the total number of historical data to be decided by the users;
$\Omega$ is a positive definite matrix;
$\varrho_{h}$ is the triggered weight of each historical data and $\sum_{h=1}^{\kappa} \varrho_{h} = 1$.

\textbf{Remark 1.} \label{remark 1}
In this paper, $h$ is the serial number of historical data, if $h=1$, the MDETM will reduces to the normal memoryless DETM.
Further more, make $x(t)=x(0)$ if $t<\kappa$.
In this paper, $\varrho_{h}$ denotes the triggered weights of historical states in the event condition. It is widely regarded the new information is more important than other old ones. Therefore, $\varrho_{1}$ should be set larger than the others and $\varrho_{h}\geq\varrho_{h+1}$ for $h=1,\cdots,\kappa$.

%C.It should be rewrite.

On the other hand, throughout all this paper, it is assumed that the state data of control plant (\ref{q1}) is transmitted to proposed controller over the network as a stream of packets, where the network-induced fading phenomenon might happen.
On account of fading channel, the common phenomenon in unreliable communication network, the actual measured output can be described by \cite{Z.Zhang2021}
\begin{equation}\label{Pre4}
y_{\xi}\left( t_k - h + 1 \right)=\xi\left(t_k\right) y\left( t_k - h + 1 \right),  \forall h=1,\cdots,\kappa
\end{equation}
with the stochastic process $\xi\left(t_k\right)$ with mathematical expectation $\bar{\xi}$ and variance $\xi^{\ast}$. %When $\xi\left(t\right) \in\{0,1\}$ , the fading model will reduce to the packet dropout case.
%Note that if the stochastic variable $\xi$ in (\ref{q4}) satisfies the Bernoulli distribution $\xi\in\{0,\ 1\}$, the fading model reduces to a data packet dropout one.
During the interval $\left[t_k,\ t_k+1\right)$, at the controller side, only $y_{\xi}\left(t_k\right) , \cdots , y_{\xi}\left(t_k - h + 1 \right),\cdots,y_{\xi}\left(t_k - \kappa + 1 \right)$ can be utilized under such fading channel model.

\subsection{MOF IT2 T-S fuzzy Controller Design:}

According to the above discussions, a type of propose $h$-order MOF controller can be described by
\begin{align}
&\textbf{Controller Rule } \mathscr{C}^{j}\textbf{: }\textbf{IF } \upsilon_{1}(y_{\xi}(t)) \textbf{ is } N_{1}^{j} \textbf{ and } \upsilon_{2}(y_{\xi}(t))) \textbf{ is }\nonumber\\
& N_{2}^{j} \textbf{ and } \cdots \textbf{ and } \upsilon_{\phi}(y_{\xi}(t)) \textbf{ is } {N}_{\phi}^{j}, \textbf{ THEN } \nonumber\\
&\ \ \ \ \ \ \ \ \ \ \ \ \ \ \ \ \ \ u(t)=\sum_{h=1}^{\kappa}K_{j}^{(h)} y_{\xi}\left(t-h+1\right)\label{q5}
\end{align}
where $\mathscr{C}^{j}$ denotes the $j$-th controller rule with $j\in\mathscr J=\left\{1,2,\cdot\cdot\cdot,\mathfrak{q}\right\}$, and $\mathfrak{q}$ symbolizes the number
of IF-THEN rules;
$\upsilon(y_{\xi}(t))=\left[\upsilon_{1}(y_{\xi}(t)), \upsilon_{2}(y_{\xi}(t)), \ldots, \upsilon_{\phi}(y_{\xi}(t))\right]$ are premise variables of the controller;
$N_{\tau}^{j}(\tau=1,2, \ldots, \phi)$ are type-2 fuzzy sets;
$K_{j}^{(h)}$ is the $h$-th local sub-controller gain of the $j$-th fuzzy rule.

The firing strength of the $j$-th rule is of the following interval sets:
\begin{align}
\mathscr{N}_{j}=\left[
    \begin{array}{c c}
        \underline{n}_{j}(y_{\xi}(t)) \quad \bar{n}_{j}(y_{\xi}(t))
    \end{array}
\right],\ j \in \mathscr{J}
\end{align}
where
\begin{equation}
\begin{gathered}
\underline{n}_{j}(y_{\xi}(t))=\prod_{\tau=1}^{\phi} \underline{\sigma}_{N_{\tau}^{j}(\upsilon(y_{\xi}(t))} \geq 0,\ \\
\bar{n}_{j}(y_{\xi}(t))=\prod_{\tau=1}^{\phi} \bar{\sigma}_{N_{\tau}^{j}(\upsilon(y_{\xi}(t)))} \geq 0, \\
\bar{\sigma}_{N_{\tau}^{j}(\upsilon(y_{\xi}(t)))} \geq \underline{\sigma}_{N_{\tau}^{j}(\upsilon(y_{\xi}(t)))} \geq 0,\ \\
\bar{n}_{j}(y_{\xi}(t)) \geq \underline{n}_{j}(y_{\xi}(t)) \geq 0
\end{gathered}
\end{equation}
with $\underline{n}_{j}(y_{\xi}(t))$ and $\bar{n}_{j}(y_{\xi}(t))$ being lower and upper membership functions, respectively; $\underline{\sigma}_{N_{\tau}^{j}(\upsilon(y_{\xi}(t)))}$ and $\bar{\sigma}_{N_{\tau}^{j}(\upsilon(y_{\xi}(t)))}$ being the lower and upper grades of the membership of $\upsilon_{\tau}(y_{\xi}(t))$ in $N_{\tau}^{j}$, respectively.
%The global fuzzy controller can be inferred as follows:
%\begin{equation}
%u(t)=\sum_{s \in \mathscr{Q}} m_{s}(y_{\xi}(t))\sum_{\kappa=1}^{h}K_{s}^{\kappa} y_{\xi}\left(t-\kappa+1\right)
%\end{equation}
%where
Let $n_{j}(y_{\xi}(t))$ represent the normalized membership function satisfying
\begin{equation}
\begin{aligned}
&n_{j}(y_{\xi}(t)) \\ &= \frac{\underline{\iota}_{j}(y_{\xi}(t)) \underline n_{j}(y_{\xi}(t))+\bar{\iota}_{j}(y_{\xi}(t)) \bar n_{j}(y_{\xi}(t))}{\sum_{k \in \mathscr{J}}\left(\underline{\iota}_{k}(y_{\xi}(t)) \underline{n}_{k}(y_{\xi}(t))+\bar{\iota}_{k}(y_{\xi}(t)) \bar{n}_{k}(y_{\xi}(t))\right)}, \\
&\ \ \ \ \ \ \ \ \ \ \ \ \ n_{j}(y_{\xi}(t)) \geq 0,\ \sum_{j \in \mathscr{J}} n_{j}(y_{\xi}(t))=1
\end{aligned}
\end{equation}
with the nonlinear functions $\underline{\iota}_{j}(y_{\xi}(t))$, $\bar{\iota}_{j}(y_{\xi}(t))$ satisfying
\begin{equation}
\begin{aligned}
0 \leq \underline{\iota}_{j}(y_{\xi}(t)), \bar{\iota}_{j}(y_{\xi}(t)) \leq 1,\\
\underline{\iota}_{j}(y_{\xi}(t))+\bar{\iota}_{j}(y_{\xi}(t))=1.
\end{aligned}
\end{equation}

%Note that the parameter uncertainties of a general nonlinear system can be described by $\underline{\beta}_{s}(y_{\xi}(t))$ and $\bar{\beta}_{s}(y_{\xi}(t))$ ,which are not necessarily known but exist and satisfy (\ref{q6}).

%Utilizing a center-average defuzzifier, product inference, and singleton fuzzifier, yields the following global memory fuzzy controller
The global fuzzy controller can be inferred as follows:
\begin{equation}\label{Pre5}
u(t)= \sum_{h=1}^{\kappa} \sum_{j \in \mathscr{J}} n_{j}(y_{\xi}(t)) K_{j}^{(h)} y_{\xi}\left(t-h+1\right).
\end{equation}

\subsection{Actuator Failure Model:}

In practical applications, due to some internal or external reasons, the actuator may suffer some faults and work inexactly. Therefore, the actuator failure model \cite{Yao2012} is considered in this paper, and the specific model can be shown as follows
\begin{equation}\label{Pre6}
u_{f}(t)=\alpha_f u(t)
\end{equation}
where $\alpha_f \in [0,1]$ denotes the failure parameter, which is a constant scalar.
Note that the relation between $\alpha_f$ and actuator failure can be expressed as follows
$$
\begin{aligned}
\left\{
\begin{array}{ll}
\alpha_f = 0, & \text{if the actuator is completely faulty};\\
\alpha_f \in (0,1), & \text{if the actuator has partial failure};\\
\alpha_f = 1, & \text{if the actuator can work normally}.
\end{array}
\right.
\end{aligned}
$$

\subsection{Closed-loop System:}

Then, combine (\ref{Pre1}), (\ref{Pre2}), (\ref{Pre4}), (\ref{Pre5}) with (\ref{Pre6}) to obtain the closed-loop
system, which can be formulated by

%\begin{equation}
%\left\{\begin{aligned}
%& x(t+1)=\mathbf{A}(w) x(t) + \mathbf{B_{u}}(w) \sum_{\kappa=1}^{h} %\mathbf{K^{(\kappa)}}(m) \xi(t-\kappa+1) \mathbf{C_y}(w) x(t-\kappa+1) + \mathbf{B_{d}}(w) %d(t)\\%\times
%& z(t)=\mathbf{C_{z}}(w) x(t).
%\end{equation}

%where $h$ is the number of the memory data, $x(t-\kappa+1)$ stands for the last $\kappa$-th history triggered data. For the purpose of facilitating the controller design, rewrite the closed loop system (\ref{q8}) as the following form
\begin{equation}\left\{\begin{aligned}\label{Pre7}
& \tilde{x}(t+1)=\tilde{A}(m,n,m,t_k)\tilde{x}(t) -\tilde{B}_{\varepsilon}(m,n,m,t_k)\\
& \ \ \ \ \ \ \ \ \ \ \ \ \ \times\tilde{\varepsilon}(t)+\tilde{B}_{d}(m)d(t)\\
& z(t)=\tilde{C}_{z}(m) \tilde{x}(t)
\end{aligned}\right.\end{equation}
with
\begin{equation*}\left\{\begin{aligned}
&\tilde{x}(t)=\left[
        x^{T}(t-\kappa+1)\ \cdots \ x^{T}(t-h+1) \  \cdots \right.\\& \ \ \ \ \ \ \ \ \ \left. x^{T}(t-1) \ x^{T}(t)
%    \begin{array}{c c c c c c}
%        x^{T}(t-\kappa+1)& \cdots & x^{T}(t-h+1)&  \cdots &x^{T}(t-1)&x^{T}(t)
%    \end{array}
\right]^{T},\\
& \tilde{A}(m,n,m,t_k)=\left[
    \begin{array}{cc}
        \begin{array}{c c}
            0_{(\kappa-1)n_x,n_x}& I_{(\kappa-1)n_x}
        \end{array}\\
        \Lambda_A(m,n,m,t_k)
    \end{array}\right],\\
& \tilde{B}_\varepsilon(m,n,m,t_k)=\left[
    \begin{array}{c}
        \begin{array}{cc}
            0_{\kappa n_x,n_x}
        \end{array}\\
        \Lambda_B(m,n,m,t_k)
    \end{array}\right],\\
& \tilde{B}_d(m)=\sum_{i \in \mathscr{I}} m_{i}(x(t)) \mathtt{\tilde{B}}_{\mathtt{d}i},\\
& \tilde{C}_z(m)=\sum_{i \in \mathscr{I}} m_{i}(x(t)) \mathtt{\tilde{C}}_{\mathtt{z}i},\\
\end{aligned}\right.\end{equation*}
\begin{equation}\left\{\begin{aligned}
& \Lambda_A(m,n,m,t_k)\\
& \ \ \ \ \ \ \ \ =\left[
    \begin{array}c
    \left(\xi\left(t_k\right)\Gamma_{A}^{(\kappa)}(m,n,m)\right)^{T}\\
    \vdots\\
    \left(\xi\left(t_k\right)\Gamma_{A}^{(h)}(m,n,m)\right)^{T}\\
    \vdots\\
    \left(\xi\left(t_k\right)\Gamma_{A}^{(2)}(m,n,m)\right)^{T}\\
    \left(\mathtt{A}(m)+\xi\left(t_k\right)\Gamma_{A}^{(1)}(m,n,m)\right)^{T}
    \end{array}\right]^{T},\\
& \Lambda_B(m,n,m,t_k)\\
& \ \ \ \ \ \ \ \ \ \ \ \ \ \ =\left[
    \begin{array}c
    \left(\xi\left(t_k\right)\Gamma_{B}^{(\kappa)}(m,n,m)\right)^{T}\\
    \vdots\\
    \left(\xi\left(t_k\right)\Gamma_{B}^{(h)}(m,n,m)\right)^{T}\\
    \vdots\\
    \left(\xi\left(t_k\right)\Gamma_{B}^{(2)}(m,n,m)\right)^{T}\\
    \left(\xi\left(t_k\right)\Gamma_{B}^{(1)}(m,n,m)\right)^{T}
    \end{array}\right]^{T},\\
& \Gamma_{A}^{(h)}(m,n,m) = \sum_{i \in \mathscr{I}} \sum_{j \in \mathscr{J}} \sum_{k \in \mathscr{I}} m_{i}(x(t)) n_{j}(y_{\xi}(t)) \\
& \ \ \ \ \ \ \ \ \ \ \ \ \ \ \ \ \ \ \ \times m_{k}(x(t)) \alpha_{f} {B_u}_{i} K_{j}^{(h)} {C_y}_{k},\\
& \Gamma_{B}^{(h)}(m,n,m) = \sum_{i \in \mathscr{I}} \sum_{j \in \mathscr{J}} \sum_{k \in \mathscr{I}} m_{i}(x(t)) n_{j}(y_{\xi}(t)) \\
& \ \ \ \ \ \ \ \ \ \ \ \ \ \ \ \ \ \ \ \times m_{k}(x(t)) \alpha_{f} {B_u}_{i} K_{j}^{(h)},\\
& \mathtt{\tilde{B}}_{\mathtt{d}ijk} = \left[
    \begin{array}{c c}
        0_{(\kappa-1)n_x, n_d}\\ {B_d}_{i}
    \end{array}\right],\\
&\mathtt{\tilde{C}}_{\mathtt{z}i} =
\left[
    \begin{array}{c c}
        0_{n_z,(\kappa-1)n_x}& {C_z}_{i}
    \end{array}\right],\\
& \mathtt{A}(m) = \sum_{i \in \mathscr{I}} m_{i}(x(t)) A_{i}.
\end{aligned}\right.\end{equation}
Later in this article, rewrite $\tilde{A}(m,n,m,t_k)$, $\tilde{B}_{\varepsilon}(m,n,m,t_k)$, $\tilde{B}_{d}(m)$, $\tilde{C}_{z}(m)$, and $\Gamma^{(h)}(m,n,m)$ as $\tilde{A}$, $\tilde{B}_{\varepsilon}$, $\tilde{B}_{d}$, $\tilde{C}_{z}$, and $\Gamma^{(h)}$ for short.

%[Noval32]
\textbf{Definition 1.} \label{definition 1} For a given disturbance attenuation level $\gamma>0$, and a scalar $\hslash \in(0,1)$ with any $c > 0$, the closed-loop system (\ref{Pre7}) is mean-square exponentially stable in the case of $d(t) = 0$ such that $\mathbb{E}\{ {\|x_{h}(t)\|}^{2}\} \leq c \hslash^{(t-t_{0})}\sup_{t_0 \in \mathscr{T}}\mathbb{E}\{\| x_{h}(t_{0}) \|^{2}\}, \ \forall t\geq t_{0}$.
the closed-loop system (\ref{Pre7}) is mean-square exponentially stable with $\mathscr H_{\infty}$ performance $\gamma$, such that
$\mathbb{E}\{ \sum_{t=0}^{\infty} \|z(t)\|^{2} \} < \gamma^{2} \sum_{t=0}^{\infty}\|d(t)\|^{2} $, holds for $d(t) \neq 0$ under zero-initial condition.

\section{Main Results}
In this section, the the mean-square exponentially stability with $\mathscr H_{\infty}$ performance of closed-loop system (\ref{Pre7}) will be exploited.

\textbf{Theorem 1.} \label{theorem 1}
Given scalars $0<\rho<1$, $\nu>0$, $0<\mu<1$, $\kappa$, $\tilde{\varrho}$, $\bar{\xi}, \xi^{*}$, $\alpha_f \in [0,1]$, $\hslash \in(0,1)$, $0<\hbar<1$, and matrix $E$, $F$, the closed-loop system (\ref{Pre7}) is mean-square exponentially stable with $\mathscr H_{\infty}$ performance index $\gamma$, if there exist scalar $\gamma>0$, $\delta>0$, symmetric matrix variables $\tilde{P}_{11}>0$, $\tilde{P}_{22}>0$, $\zeta_{j}$, $\Omega$, and matrix variables $\tilde{P}_{12}={\tilde{P}_{21}}^{T}$, $\eta^{(h)}_{j}$ satisfying

\begin{equation}\label{Main0}
%\begin{aligned}
\tilde{P}>0,
%\end{aligned}
\end{equation}
\begin{equation}\label{Main1}
%\begin{aligned}
\Psi_{ijk} < 0,\ \forall i, j, k
%\end{aligned}
\end{equation}
with

\begin{equation}
\left\{ \begin{aligned}
%\begin{array}{llll}
    &\Psi_{ijk} =
        \left[ \begin{array}{cccccc}
        \Phi_{ik}                &\ast                                      &\ast                                                 \\
        0                                &-\gamma^{2}I_{n_d}                        &\ast                                           \\
        0                                &0                                         &\frac{\mu-(1-\hslash)+\delta}{\nu}I_{n_\varpi} \\
        \bar{\Sigma}_{ijk}               &\tilde{P}\mathtt{\tilde{B}}_{\mathtt{d}i} &0                                              \\
        {\Sigma}^{\ast}_{ijk}            &0                                         &0                                             \\
        \theta\Xi_{jk}                   &0                                         &0
        \end{array} \right.\\
&\ \ \ \ \ \ \ \ \ \left. \begin{array}{cccccc}
        &\ast                 &\ast                    &\ast \\
        &\ast                 &\ast                    &\ast \\
        &\ast                 &\ast                    &\ast \\
        &-{\tilde{P}}         &0                       &\ast \\
        &0                    &-{\tilde{P}}            &\ast \\
        &{\bar{\Xi}}^{T}_{ij} &{{\Xi}^{\ast}}^{T}_{ij} &-\theta\zeta_{j}-\theta{\zeta}^{T}_{j}
        \end{array} \right],\\
&\Phi_{ik}=
    \left[\begin{array}{cc}
    \Phi_{ik}^{(1)}
    &\ast \\
    0                     &-(\frac{1}{\nu}+\delta){\tilde{\varrho} }^{T}\Omega\tilde{\varrho} \end{array}\right],\\
&\Phi_{ik}^{(1)}=
        -(1-\hslash)\tilde{P}+(\frac{1}{\nu}+\delta)\rho H^T{C_y}_{k}^{T}\Omega{C_y}_{k}H \\ & \ \ \ \ \ \ \ \ +\mathtt{\tilde{C}}_{\mathtt{z}i}^{T}\mathtt{\tilde{C}}_{\mathtt{z}i},\\
& \bar{\Sigma}_{ijk}=
      \left[ \begin{array}{cc}
        \bar{\Sigma}_{ijk}^{(1)}   & \bar{\Sigma}_{ijk}^{(2)}
     \end{array} \right],\\
&\bar{\Sigma}_{ijk}^{(1)}=\triangle
       +\left[ \begin{array}{cccccc}
            \bar{\xi}{\Upsilon^{(\kappa)}_{ijk}}^{T}\\
            \cdots\\
            \bar{\xi}{\Upsilon^{(h)}_{ijk}}^{T}\\
            \cdots\\
            A^{T}_{i}
            \left[ \begin{array}{cc}
                \tilde{P}_{12}\\
                \tilde{P}_{22}
            \end{array} \right]^{T} +
            \bar{\xi}{\Upsilon^{(1)}_{ijk}}^{T}
        \end{array} \right]^{T},\\
&\bar{\Sigma}_{ijk}^{(2)}=-\left[ \begin{array}{cccccc}
            \bar{\xi}{\Upsilon^{(\kappa)}_{ijk}} &
            \cdots &
            \bar{\xi}{\Upsilon^{(h)}_{ijk}} &
            \cdots &
            \bar{\xi}{\Upsilon^{(1)}_{ijk}}
        \end{array} \right],\\
&\triangle=\left[ \begin{array}{cc}
            \tilde{P}_{11}\\
            \tilde{P}_{21}
        \end{array} \right]
        \left[ \begin{array}{cc}
            0_{n_x,(\kappa-1)n_x}& I_{(\kappa-1)n_x}
        \end{array} \right],\\
& {\Upsilon}^{(h)}_{ijk} = {\alpha}_{f}\left[ \begin{array}{cc} E \\ F \end{array} \right]{B_u}_{i}\eta^{(h)}_{j}{C_y}_{k},\\
& \bar{\Xi}_{ij}=\left[
    \begin{array}{cc}
        \bar{\xi}{\alpha}_{f}(\tilde{P}_{12}{B_u}_{i} - E{B_u}_{i}\zeta_{j})\\
        \bar{\xi}{\alpha}_{f}(\tilde{P}_{22}{B_u}_{i} - F{B_u}_{i}\zeta_{j})
    \end{array}\right],\\
& \Xi_{jk}=\left[
        \eta^{(\kappa)}_{j}{C_y}_{k} \ \cdots \ \eta^{(h)}_{j}{C_y}_{k} \ \cdots \ \eta^{(1)}_{j}{C_y}_{k} \right.\\&\left. \ \ \ \ \ \ \
        \ -\eta^{(\kappa)}_{j}{C_y}_{k} \ \cdots \ -\eta^{(h)}_{j}{C_y}_{k} \ \cdots \ -\eta^{(1)}_{j}{C_y}_{k}\right],\\
& {\Sigma}^{\ast}_{ijk}= \left[ \begin{array}{cc}
    {\Sigma}^{\ast(1)}_{ijk}
    &{\Sigma}^{\ast(2)}_{ijk}
    \end{array} \right],\\
&{\Sigma}^{\ast(1)}_{ijk}=\left[
        \sqrt{{\xi}^{\ast}}{\Upsilon^{(\kappa)}_{ijk}}
        \cdots
        \sqrt{{\xi}^{\ast}}{\Upsilon^{(h)}_{ijk}}
        \cdots
        \sqrt{{\xi}^{\ast}}{\Upsilon^{(1)}_{ijk}}
    \right],\\
&{\Sigma}^{\ast(2)}_{ijk}=-\left[
        \sqrt{{\xi}^{\ast}}{\Upsilon^{(\kappa)}_{ijk}}
        \cdots
        \sqrt{{\xi}^{\ast}}{\Upsilon^{(h)}_{ijk}}
        \cdots
        \sqrt{{\xi}^{\ast}}{\Upsilon^{(1)}_{ijk}}
    \right],\\
& {\Xi}^{\ast}_{ij}=\left[
    \begin{array}{cc}
        \sqrt{{\xi}^{\ast}}{\alpha}_{f}(\tilde{P}_{12}{B_u}_{i} - E{B_u}_{i}\zeta_{j})\\
        \sqrt{{\xi}^{\ast}}{\alpha}_{f}(\tilde{P}_{22}{B_u}_{i} - F{B_u}_{i}\zeta_{j})
    \end{array}\right].
%\end{array}
\end{aligned}\right.
\end{equation}

%for any $l=1,2, \ldots, p, s=$ $1,2, \ldots, q$,
Moreover, the control gains are given by $ K^{(h)}_{j} = {\zeta}^{-1}_{j} \eta^{(h)}_{j} $.

\textbf{Proof.}
%Now, construct the Lyapunov function as
Consider the construction of the Lyapunov function as follows:
\begin{equation}
V(t)=\tilde{x}^{T}(t) \tilde{P} \tilde{x}(t) + \frac{1}{\nu} \varpi(t)
\end{equation}\label{Main8}
where $\tilde{P}=\left [ \begin{array}{cc} \tilde{P}_{11} &  \tilde{P}_{12}\\ \tilde{P}_{21} &  \tilde{P}_{22} \end{array}\right]\in {\Re^{\kappa n_{x}\times \kappa n_{x}}}$, and
$\tilde{P}_{11}\in{\Re^{(\kappa-1)n_{x}\times (\kappa-1)n_{x}}}$,
$\tilde{P}_{12}\in{\Re^{(\kappa-1)n_{x}\times \kappa n_{x}}}$,
$\tilde{P}_{21}\in{\Re^{\kappa n_{x}\times (\kappa-1)n_{x}}}$,
$\tilde{P}_{22}\in{\Re^{n_{x}\times n_{x}}}$.

Note that for the random variable $\xi(t_k)$, $E\{ (\bar{\xi}-\xi(t_k))^{2} \} = E\{ {\bar{\xi}}^{2} - 2\bar{\xi}\xi(t_k) + {\xi}^{2}(t_k)\} = {\xi}^{\ast}$ due to $E\{ {\xi}^{2}(t_k)\} = {\xi}^{\ast} + {\bar{\xi}}^{2}$.
Consequently, one has
\begin{equation}\begin{aligned}\label{Main6}
\mathbb{E}\{\tilde{x}^{T}&(t+1) \tilde{P} \tilde{x}(t+1) \} \\
=& \mathbb{E}\{ (\tilde{A}\tilde{x}(t)-\tilde{B}_{\varepsilon}\tilde{\varepsilon}(t)+\tilde{B}_{d}d(t))^{T} \tilde{P} (\tilde{A}\tilde{x}(t)-\tilde{B}_{\varepsilon}\tilde{\varepsilon}(t) \\ & +\tilde{B}_{d}d(t)) \} \\
=& (\bar{\tilde{A}}\tilde{x}(t)-\bar{\tilde{B}}_{\varepsilon}\tilde{\varepsilon}(t)+\tilde{B}_{d}d(t))^{T} \tilde{P} (\bar{\tilde{A}}\tilde{x}(t)-\bar{\tilde{B}}_{\varepsilon}\tilde{\varepsilon}(t) \\ & +\tilde{B}_{d}d(t)) + (\tilde{A}^{\ast}\tilde{x}(t)-{\tilde{B}_{\varepsilon}}^{\ast}\tilde{\varepsilon}(t))^{T} \tilde{P} (\tilde{A}^{\ast}\tilde{x}(t) \\ & -{\tilde{B}_{\varepsilon}}^{\ast}\tilde{\varepsilon}(t))
\end{aligned}\end{equation}

where
\begin{equation*}\left\{\begin{aligned}
& \bar{\tilde{A}}(m,n,m)=\left[
    \begin{array}{cc}
        \begin{array}{c c}
            0_{(\kappa-1)n_x,n_x}& I_{(\kappa-1)n_x}
        \end{array}\\
        \bar{\Lambda}_A(m,n,m)
    \end{array}\right],\\
& \bar{\tilde{B}}_{\varepsilon}(m,n,m)=\left[
    \begin{array}{c}
        \begin{array}{cc}
            0_{\kappa n_x,n_x}
        \end{array}\\
        \bar{\Lambda}_B(m,n,m)
    \end{array}\right],\\
& \tilde{A}^{\ast}(m,n,m)=\left[
    \begin{array}{cc}
        \begin{array}{c c}
            0_{\kappa n_x,n_x}
        \end{array}\\
        {\Lambda}^{\ast}_{A}(m,n,m)
    \end{array}\right],\\
& {\tilde{B}_{\varepsilon}}^{\ast}(m,n,m)=\left[
    \begin{array}{c}
        \begin{array}{cc}
            0_{\kappa n_x,n_x}
        \end{array}\\
        {\Lambda}^{\ast}_{B}(m,n,m)
    \end{array}\right],\\
& \bar{\Lambda}_A(m,n,m)\\
&\ \ \ \ \ \ \ \ \ \ =\left[
    \begin{array}c
    \left(\bar{\xi}\Gamma_{A}^{(\kappa)}(m,n,m)\right)^{T}\\
    \vdots\\
    \left(\bar{\xi}\Gamma_{A}^{(h)}(m,n,m)\right)^{T}\\
    \vdots\\
    \left(\bar{\xi}\Gamma_{A}^{(2)}(m,n,m)\right)^{T}\\
    \left(\mathtt{A}(m)+\bar{\xi}\Gamma_{A}^{(1)}(m,n,m)\right)^{T}
    \end{array}\right]^{T},\\
& \bar{\Lambda}_B(m,n,m)=\left[
    \begin{array}c
    \left(\bar{\xi}\Gamma_{B}^{(\kappa)}(m,n,m)\right)^{T}\\
    \vdots\\
    \left(\bar{\xi}\Gamma_{B}^{(h)}(m,n,m)\right)^{T}\\
    \vdots\\
    \left(\bar{\xi}\Gamma_{B}^{(2)}(m,n,m)\right)^{T}\\
    \left(\bar{\xi}\Gamma_{B}^{(1)}(m,n,m)\right)^{T}
    \end{array}\right]^{T},\\
\end{aligned}\right.\end{equation*}
\begin{equation}\left\{\begin{aligned}
& {\Lambda}^{\ast}_{A}(m,n,m)=\left[
    \begin{array}c
    \left(\sqrt{{\xi}^{\ast}} \Gamma_{A}^{(\kappa)}(m,n,m)\right)^{T}\\
    \vdots\\
    \left(\sqrt{{\xi}^{\ast}} \Gamma_{A}^{(h)}(m,n,m)\right)^{T}\\
    \vdots\\
    \left(\sqrt{{\xi}^{\ast}} \Gamma_{A}^{(2)}(m,n,m)\right)^{T}\\
    \left(\sqrt{{\xi}^{\ast}} \Gamma_{A}^{(1)}(m,n,m)\right)^{T}
    \end{array}\right]^{T},\\
& {\Lambda}^{\ast}_{B}(m,n,m)=\left[
    \begin{array}c
    \left(\sqrt{{\xi}^{\ast}} \Gamma_{B}^{(\kappa)}(m,n,m)\right)^{T}\\
    \vdots\\
    \left(\sqrt{{\xi}^{\ast}} \Gamma_{B}^{(h)}(m,n,m)\right)^{T}\\
    \vdots\\
    \left(\sqrt{{\xi}^{\ast}} \Gamma_{B}^{(2)}(m,n,m)\right)^{T}\\
    \left(\sqrt{{\xi}^{\ast}} \Gamma_{B}^{(1)}(m,n,m)\right)^{T}
    \end{array}\right]^{T}.
\end{aligned}\right.\end{equation}

Later in this article, rewrite $\bar{\tilde{A}}(m,n,m)$, $\bar{\tilde{B}}_{\varepsilon}(m,n,m)$, $\tilde{A}^{\ast}(m,n,m)$, ${\tilde{B}_{\varepsilon}}^{\ast}(m,n,m)$, $\bar{\Lambda}_A(m,n,m)$, $\bar{\Lambda}_B(m,n,m)$, ${\Lambda}^{\ast}_{A}(m,n,m)$, and ${\Lambda}^{\ast}_{B}(m,n,m)$ as $\bar{\tilde{A}}$, $\bar{\tilde{B}}_{\varepsilon}$, $\tilde{A}^{\ast}$, ${\tilde{B}_{\varepsilon}}^{\ast}$, $\bar{\Lambda}_A$, $\bar{\Lambda}_B$, ${\Lambda}^{\ast}_{A}$, and ${\Lambda}^{\ast}_{B}$ for short.

On the other hand, by introducing the DETM condition (\ref{Pre2}), we obtain for any $k \in [t_{k}, t_{k+1})$
\begin{equation}
    \frac{1}{\nu} \varpi(t) + \rho{x}^{T}(t)\Omega x(t) -  \hat{\varepsilon}^{T}(t)\Omega\hat{\varepsilon}(t) \geq 0 ,
\end{equation}
which implies that for any $\delta > 0$
\begin{equation}\begin{aligned}\label{Main7}
\frac{1}{\nu}\varpi&(t+1)-\frac{(1-\hslash)}{\nu} \varpi(t)\\
\leq& \frac{1}{\nu} (\mu\varpi(t) + \rho{x}^{T}(t)\Omega x(t) - \hat{\varepsilon}^{T}(t)\Omega\hat{\varepsilon}(t))\\
&- \frac{(1-\hslash)}{\nu} \varpi(t) + \delta(\frac{1}{\nu} \varpi(t)+ \rho{x}^{T}(t)\Omega x(t) \\
&-  \hat{\varepsilon}^{T}(t)\Omega\hat{\varepsilon}(t))\\
=& \frac{\mu-(1-\hslash)+\delta}{\nu}\varpi(t) +(\frac{1}{\nu}+\delta)\rho{x}^{T}(t)\Omega x(t) \\
&- (\frac{1}{\nu}+\delta)\hat{\varepsilon}^{T}(t)\Omega\hat{\varepsilon}(t)\\
=& \frac{\mu-(1-\hslash)+\delta}{\nu}\varpi(t) +(\frac{1}{\nu}+\delta)\rho\tilde{x}^{T}(t) H^T \\
& \times\Omega H \tilde{x}(t)- (\frac{1}{\nu}+\delta){\tilde{\varepsilon}}^{T}(t){\tilde{\varrho} }^{T}\Omega\tilde{\varrho} \tilde{\varepsilon}(t)
\end{aligned}\end{equation}

with $H=[ \begin{array}{c c }  0_{n_x,(\kappa-1)n_x}&  I_{n_x} \end{array} ]$.

Consider about $ m_{i}(x(t)) n_{j}(y_{\xi}(t)) m_{k}(x(t)) > 0$ for any ${i \in \mathscr{I}}, {j \in \mathscr{J}}, {k \in \mathscr{I}}$, by using (\ref{Main1}),  one can get that
\begin{equation}\label{Main4}
%\begin{aligned}
\sum_{i \in \mathscr{I}} \sum_{j \in \mathscr{J}} \sum_{k \in \mathscr{I}} m_{i}(x(t)) n_{j}(y_{\xi}(t)) m_{k}(x(t))
\Psi_{ijk} < 0.
%\end{aligned}
\end{equation}

On the other hand, there defines some new notations:
\begin{equation}\label{Main2}
\begin{aligned}
    \bar{\Theta}=\left[\begin{array}{cc} \bar{\tilde{A}} &-\bar{\tilde{B}}_{\varepsilon} \end{array}\right], \
    {\Theta}^{\ast}=\left[\begin{array}{cc} \tilde{A}^{\ast} &-{\tilde{B}_{\varepsilon}}^{\ast} \end{array}\right] .
\end{aligned}
\end{equation}

Then, note that (\ref{Main2}) implies that
%\begin{equation*}
%\begin{aligned}
%\tilde{P}{\bar{\Theta}}
%=&
%\left[ \begin{array}{cc}
% \tilde{P}\bar{\tilde{A}} &-\tilde{P}\bar{\tilde{B}}_{\varepsilon}
%\end{array} \right]\\
%=& \left[ \begin{array}{cc}
%    \Delta
%    +
%    \left[ \begin{array}{cc}
%        \tilde{P}_{12}\\
%        \tilde{P}_{22}
%    \end{array} \right]
%    \bar{\Lambda}_A
%    &-
%    \left[ \begin{array}{cc}
%        \tilde{P}_{12}\\
%        \tilde{P}_{22}
%    \end{array} \right]
%    \bar{\Lambda}_B
%\end{array} \right]\\
%=& \sum_{i \in \mathscr{I}} \sum_{j \in \mathscr{J}} \sum_{k \in \mathscr{I}} m_{i}(x(t)) n_{j}(y_{\xi}(t)) m_{k}(x(t))\\
%&  \times \left( \bar{\Sigma}_{ijk} + \bar{\Xi}_{ij}{\zeta}^{-1}_{j}\Xi_{jk} \right),
%\end{aligned}
%\end{equation*}
\begin{equation}\label{Main3}
\begin{aligned}
\tilde{P}{\bar{\Theta}}
=&
\left[ \begin{array}{cc}
 \tilde{P}\bar{\tilde{A}} &-\tilde{P}\bar{\tilde{B}}_{\varepsilon}
\end{array} \right]\\
=& \left[ \begin{array}{cc}
    \Delta
    +
    \left[ \begin{array}{cc}
        \tilde{P}_{12}\\
        \tilde{P}_{22}
    \end{array} \right]
    \bar{\Lambda}_A
    &-
    \left[ \begin{array}{cc}
        \tilde{P}_{12}\\
        \tilde{P}_{22}
    \end{array} \right]
    \bar{\Lambda}_B
\end{array} \right]\\
=& \sum_{i \in \mathscr{I}} \sum_{j \in \mathscr{J}} \sum_{k \in \mathscr{I}} m_{i}(x(t)) n_{j}(y_{\xi}(t)) m_{k}(x(t))\\
&  \times \left( \bar{\Sigma}_{ijk} + \bar{\Xi}_{ij}{\zeta}^{-1}_{j}\Xi_{jk} \right),\\
\tilde{P}{{\Theta}^{\ast}}
=&
\left[ \begin{array}{cc}
 \tilde{P}\tilde{A}^{\ast} &-\tilde{P}{\tilde{B}_{\varepsilon}}^{\ast}
\end{array} \right]\\
=& \left[ \begin{array}{cc}
    \left[ \begin{array}{cc}
        \tilde{P}_{12}\\
        \tilde{P}_{22}
    \end{array} \right]
    {\Lambda}^{\ast}_{A}
    &-
    \left[ \begin{array}{cc}
        \tilde{P}_{12}\\
        \tilde{P}_{22}
    \end{array} \right]
    {\Lambda}^{\ast}_{B}
\end{array} \right]\\
=& \sum_{i \in \mathscr{I}} \sum_{j \in \mathscr{J}} \sum_{k \in \mathscr{I}} m_{i}(x(t)) n_{j}(y_{\xi}(t)) m_{k}(x(t))\\
   &\times\left( {\Sigma}^{\ast}_{ijk} + {\Xi}^{\ast}_{ij}{\zeta}^{-1}_{j}\Xi_{jk} \right)
\end{aligned}
\end{equation}
with
\begin{equation}
\begin{aligned}
    \Delta=\left[ \begin{array}{cc}
        \tilde{P}_{11}\\
        \tilde{P}_{21}
    \end{array} \right]
    \left[ \begin{array}{cc}
        0_{(\kappa-1)n_x,n_x} \\ I_{(\kappa-1)n_x}
    \end{array} \right]^{T}
\end{aligned}
\end{equation}

Applying (\ref{Main3}) and Lemma 2 in Appendix to (\ref{Main4}), such that
\begin{equation}\label{Main5}
%\begin{aligned}
\left[ \begin{array}{ccccc}
 \mathbf{\Phi}              &\ast                     &\ast                                &\ast         &\ast \\
 0                          &-\gamma^{2}I_{n_d}       &\ast                                &\ast         &\ast \\
 0                          &0                        &\frac{\mu-(1-\hslash)+\delta}{\nu}I &\ast         &\ast \\
 \tilde{P}{\bar{\Theta}}    &\tilde{P}{\tilde{B}_{d}} &0                                   &-{\tilde{P}} &0 \\
 \tilde{P}{{\Theta}^{\ast}} &0                        &0                                   &0            &-{\tilde{P}}
\end{array} \right] < 0
%\end{aligned}
\end{equation}
with $\mathbf{\Phi}=\sum_{i \in \mathscr{I}} \sum_{k \in \mathscr{I}} m_{i}(x(t)) m_{k}(x(t)) \Phi_{ik}$.

Pre- and post-multiply (\ref{Main5}) with $diag\{ I,I,I,\tilde{P}^{-1},\tilde{P}^{-1} \}$, one has
\begin{equation}\label{Main9}
%\begin{aligned}
\left[ \begin{array}{ccccc}
 \mathbf{\Phi}     &\ast               &\ast                                &\ast              &\ast \\
 0                 &-\gamma^{2}I_{n_d} &\ast                                &\ast              &\ast \\
 0                 &0                  &\frac{\mu-(1-\hslash)+\delta}{\nu}I &\ast              &\ast \\
 {\bar{\Theta}}    &{\tilde{B}_{d}}    &0                                   &-{\tilde{P}}^{-1} &0 \\
 {{\Theta}^{\ast}} &0                  &0                                   &0                 &-{\tilde{P}}^{-1}
\end{array} \right] < 0.
%\end{aligned}
\end{equation}

Consider (\ref{Pre7}), (\ref{Main6}), and (\ref{Main7}) into (\ref{Main8}), the mean-square exponentially stability with $\mathscr H_{\infty}$ performance $\gamma$ of the system can be formulated by\begin{equation*}
\begin{aligned}
\mathbb{E}&\{V(t+1) - V(t) + z^{T}(t)z(t) - \gamma^{2}d^{T}(t)d(t) \} \\
&+ \hslash\mathbb{E}\{V(t)\} \\
=&\mathbb{E}\{\tilde{x}^{T}(t+1) \tilde{P} \tilde{x}(t+1) \} - (1-\hslash) \tilde{x}^{T}(t) \tilde{P} \tilde{x}(t) \\
&+ \frac{1}{\nu} \varpi(t+1)-\frac{(1-\hslash)}{\nu} \varpi(t) + z^{T}(t)z(t) - \gamma^{2}d^{T}(t)d(t) \\
\leq& (\bar{\tilde{A}}\tilde{x}(t)-\bar{\tilde{B}}_{\varepsilon}\tilde{\varepsilon}(t)+\tilde{B}_{d}d(t))^{T} \tilde{P}(\bar{\tilde{A}}(w,m)\tilde{x}(t) \\
&-\bar{\tilde{B}}_{\varepsilon}\tilde{\varepsilon}(t)+\tilde{B}_{d}d(t)) \\
&+ (\tilde{A}^{\ast}\tilde{x}(t)-{\tilde{B}_{\varepsilon}}^{\ast}\tilde{\varepsilon}(t))^{T} \tilde{P} (\tilde{A}^{\ast}\tilde{x}(t)-{\tilde{B}_{\varepsilon}}^{\ast}\tilde{\varepsilon}(t)) \\
&- (1-\hslash) \tilde{x}^{T}(t) \tilde{P} \tilde{x}(t) \\
&+ \frac{\mu-(1-\hslash)+\delta}{\nu}\varpi(t) + \left(\frac{1}{\nu}+\delta\right)\rho\tilde{x}^{T}(t) H^T\Omega H \tilde{x}(t) \\
&- \left(\frac{1}{\nu}+\delta\right){\tilde{\varepsilon}}^{T}(t){\tilde{\varrho} }^{T}\Omega\tilde{\varrho} \tilde{\varepsilon}(t) + z^{T}(t)z(t) - \gamma^{2}d^{T}(t)d(t)
\end{aligned}
\end{equation*}

\begin{equation}\label{Main10}
\begin{aligned}
=& \left[
\begin{array}{ccc}
    \wp(t) \\ d(t) \\ \sqrt{\varpi(t)}
\end{array} \right]^{T}
\left[ \begin{array}{ccc}
    {\bar{\Theta}}^{T}\tilde{P}\bar{\Theta}+{{\Theta}^{\ast}}^{T}\tilde{P}{\Theta}^{\ast}+\mathbf{\Phi}  \\
     -{\tilde{B}_{d}}^{T}\tilde{P}\bar{\Theta} \\
     0
\end{array}\right.\\
&\left.
\begin{array}{ccc}
    \ast &\ast \\
    {\tilde{B}_{d}}^{T}\tilde{P}\tilde{B}_{d}-\gamma^{2}I_{n_d} &\ast \\
    0 &\frac{\mu-(1-\hslash)+\delta}{\nu}I
\end{array} \right]
\left[ \begin{array}{ccc}
    \wp(t) \\ d(t) \\ \sqrt{\varpi(t)}
\end{array} \right]
\end{aligned}
\end{equation}
with $\wp(t)=\left[\begin{array}{cc} {\tilde{x}}^{T}(t) &{\tilde{\varepsilon}}^{T}(t) \end{array}\right]^{T}$.

By using (\ref{Main10}) and applying the Schur compliment to (\ref{Main9}), the following inequality can be guaranteed:
\begin{equation}\begin{aligned}
\mathbb{E}\{V(t+1) - V(t) + z^{T}(t)z(t) - \gamma^{2}d^{T}(t)d(t) \} \\ < - \hslash \mathbb{E}\{V(t) \},
\end{aligned}\end{equation}
%which can establish sufficient conditions to achieve the mean-square exponentially stability with $\mathscr H_{\infty}$ performance $\gamma$ of the closed-loop system, by utilizing the proof of the Lemma \ref{lemma 3} in Appendix. The proof is completed.
%
%
%Then, to reduce the conservation of the result in Theorem \ref{theorem 1}, inspired by [Final 19], by exploiting the information of membership functions, a MFD controller synthesis result will be summarized in the follows.

% Establishing Sufficient Conditions for Stability and Performance

which establishes sufficient conditions to achieve mean-square exponential stability with $\mathscr{H}_{\infty}$ performance $\gamma$ for the closed-loop system, through the utilization of the proof detailed in Lemma \ref{lemma 3} in the Appendix. This completes the proof.
\hfill $\square$

% Refining Controller Synthesis to Reduce Conservatism

To reduce the conservatism inherent in the result of Theorem \ref{theorem 1}, insights from \cite{HKLam2014} are incorporated. By fully exploiting the information contained within the membership functions, a more refined MFD controller synthesis result is summarized as follows.

\textbf{Theorem 2}\label{theorem 2}
Given scalars $0<\rho<1$, $\nu>0$, $0<\mu<1$, $\kappa$, $\tilde{\varrho}$, $\bar{\xi}, \xi^{*}$, $\alpha_f \in [0,1]$, $\hslash \in(0,1)$, $0<\hbar<1$, $\underline{o}_{ijk i_{1} i_{2} \cdots i_{n_{x}} j_{1} j_{2} \cdots j_{n_{y}} ls\ell}$, $\bar{o}_{ijk i_{1} i_{2} \cdots i_{n_{x}} j_{1} j_{2} \cdots j_{n_{y}} ls\ell}$ and matrix $E$, $F$, the closed-loop system (\ref{Pre7}) is mean-square exponentially stable with $\mathscr H_{\infty}$ performance index $\gamma$, if there exist scalar $\gamma>0$, $\delta>0$, symmetric matrix variables $\tilde{P}_{11}>0$, $\tilde{P}_{22}>0$, $\zeta_{j}$, $\Omega$, $M_{ijk\ell}$, $W$ and matrix variables $\tilde{P}_{12}={\tilde{P}_{21}}^{T}$, $\eta^{(h)}_{j}$ satisfying
\begin{equation}\label{MainT27}
%\begin{aligned}
\tilde{P}>0,
%\end{aligned}
\end{equation}
\begin{equation}\begin{aligned}\label{MainT25}
& \sum_{i \in \mathscr{I}} \sum_{j \in \mathscr{J}} \sum_{k \in \mathscr{I}}
    \left( \underline{o}_{ijk i_{1} i_{2} \cdots i_{n_{x}} j_{1} j_{2} \cdots j_{n_{y}} ls\ell}
        \Psi_{ijk}\right.\\
&\left.        +
        \left(
            \bar{o}_{ijk i_{1} i_{2} \cdots i_{n_{x}} j_{1} j_{2} \cdots j_{n_{y}} ls\ell}
            -
            \underline{o}_{ijk i_{1} i_{2} \cdots i_{n_{x}} j_{1} j_{2} \cdots j_{n_{y}} ls\ell}
        \right) \right. \\
&\left.        M_{ijk\ell}
        +
        \underline{o}_{ijk i_{1} i_{2} \cdots i_{n_{x}} j_{1} j_{2} \cdots j_{n_{y}} ls\ell}
        W
    \right)
    - W < 0 , \\ &\forall i_{1}, i_{2}, \cdots, i_{n_{x}}, j_{1}, j_{2}, \cdots, j_{n_{y}}, l, s, \ell, \\
\end{aligned}\end{equation}
\begin{equation}\begin{aligned}\label{MainT26}
\Psi_{ijk} - M_{ijk\ell}& + W  < 0 , \ \forall i, j, k.
\end{aligned}\end{equation}

\textbf{Proof.}
%Based on the proof of Theorem 2, if the inequality (39) can be shown, the claimed results in Theorem 2 will follow. In the sequel,  with the inequality (39), less conservative membership-function-dependent controller design results in Theorem 3 will be proven.
Inspired by [Final 19]\cite{H.K.Lam2014} and \cite{M.Wang2021}, the state spaces of the plant and controller, i.e., $x(t)$ and $y_{\xi}(t)$ are divided into $p$, $q$ connected subspaces, respectively.
In addition, the footprint of uncertainty (FOU) is divided into $\wp$ sub-FOUs.
%In the $\ell$th sub-FOU, $\ell = 1, 2, \cdots, \wp$.

Furthermore, defining $w_{ijk}(x(t),y_{\xi}(t)) = m_{i}(x(t))n_{j}(y_{\xi}(t))m_{k}(x(t))$. Then, the lower and upper membership functions $\underline{w}_{ijk\ell}(x(t),y_{\xi}(t))$, $\bar{w}_{ijk\ell}(x(t),y_{\xi}(t))$ can be defined as
\begin{equation}\begin{gathered}
\begin{aligned}
\underline{w}_{ijk\ell}&(x(t),y_{\xi}(t)) \\ = & \sum_{l=1}^{p} \sum_{s=1}^{q}
    \sum_{i_{1}=1}^{2} \cdots \sum_{i_{n_{x}}=1}^{2} \sum_{j_{1}=1}^{2} \cdots \sum_{j_{n_{y}}=1}^{2}  \prod_{\imath=1}^{n_{x}} \prod_{\jmath=1}^{n_{y}} \\
    &v_{\imath\jmath i_{\imath} j_{\jmath} ls\ell}\left(x_{\imath}(t),{y_{\xi}}_{\jmath}(t)\right)\underline{o}_{ijk i_{1} i_{2} \cdots i_{n_{x}} j_{1} j_{2} \cdots j_{n_{y}} ls\ell}, \\
&  \forall i, j, k, l, s, \ell, \\
\bar{w}_{ijk\ell}&(x(t),y_{\xi}(t)) \\ =& \sum_{l=1}^{p} \sum_{s=1}^{q}
    \sum_{i_{1}=1}^{2} \cdots \sum_{i_{n_{x}}=1}^{2} \sum_{j_{1}=1}^{2} \cdots \sum_{j_{n_{y}}=1}^{2}  \prod_{\imath=1}^{n_{x}} \prod_{\jmath=1}^{n_{y}} \\&
    v_{\imath\jmath i_{\imath} j_{\jmath} ls\ell}\left(x_{\imath}(t),{y_{\xi}}_{\jmath}(t)\right) \bar{o}_{ijk i_{1} i_{2} \cdots i_{n_{x}} j_{1} j_{2} \cdots j_{n_{y}} ls\ell}, \\
&  \forall i, j, k, l, s, \ell, \\
 \end{aligned}\\
 0 \leq \underline{w}_{ijk\ell}(x(t),y_{\xi}(t)) \leq \bar{w}_{ijk\ell}(x(t),y_{\xi}(t)), \\
  0 \leq \underline{o}_{ijk i_{1} i_{2} \cdots i_{n_{x}} j_{1} j_{2} \cdots j_{n_{y}} ls\ell} \\ \leq \bar{o}_{ijk i_{1} i_{2} \cdots i_{n_{x}} j_{1} j_{2} \cdots j_{n_{y}} ls\ell} \leq 1
\end{gathered}    \label{q26}\end{equation}
where $\underline{o}_{ijk i_{1} i_{2} \cdots i_{n_{x}} j_{1} j_{2} \cdots j_{n_{y}} ls\ell}$,
and $\bar{o}_{ijk i_{1} i_{2} \cdots i_{n_{x}} j_{1} j_{2} \cdots j_{n_{y}} ls\ell}$
are constant scalars to be determined and $v_{\imath\jmath i_{\imath} j_{\jmath} ls\ell}\left(x_{\imath}(t),{y_{\xi}}_{\jmath}(t)\right) \in [0,\ 1]$ satisfies that:
$\sum_{i_{\imath}}^{2}\sum_{j_{\jmath}}^{2} v_{\imath\jmath i_{\imath} j_{\jmath} ls\ell}\left(x_{\imath}(t),{y_{\xi}}_{\jmath}(t)\right) = 1$
if it is in the corresponding sub-value-spaces; otherwise, it's 0.
 %$v_{jk l_{j} s_{k} \imath\jmath\ell}\left(x_{j}(t),{y_{\xi}}_{k}(t)\right)$ and $v_{jk l_{j} s_{k} \imath\jmath\ell}\left(x_{j}(t),{y_{\xi}}_{k}(t)\right)$
%are the lower and upper membership functions of $h_{ls}(x(t),y_{\xi}(t))$;

Note that $\sum_{l=1}^{p} \sum_{s=1}^{q}\sum_{i_{1}=1}^{2} \cdots \sum_{i_{n_{x}}=1}^{2} \sum_{j_{1}=1}^{2} \cdots \sum_{j_{n_{y}}=1}^{2}  $\\$ \prod_{\imath=1}^{n_{x}} \prod_{\jmath=1}^{n_{y}}
    v_{\imath\jmath i_{\imath} j_{\jmath} ls\ell}\left(x_{\imath}(t),{y_{\xi}}_{\jmath}(t)\right)=1$, and $w_{ijk}(x(t),y_{\xi}(t)) = m_{i}(x(t)) n_{j}(y_{\xi}(t)) m_{k}(x(t))$, one has
\begin{equation}
\begin{aligned}
w_{ijk}&(x(t),y_{\xi}(t))\\
=& \sum_{\ell=1}^{\wp} \vartheta_{ijk\ell}(x(t),y_{\xi}(t))\left(\underline{\beta}_{ijk\ell}(x(t),y_{\xi}(t)) \right.\\ &\left. \times\underline{w}_{ijk\ell}(x(t),y_{\xi}(t))+\bar{\beta}_{ijk\ell}(x(t),y_{\xi}(t)) \right.\\
&\left. \times\bar{w}_{ijk\ell}(x(t),y_{\xi}(t))\right),\ \forall i, j, k
\end{aligned}
\end{equation}
where
$\vartheta_{ijk\ell}(x(t),y_{\xi}(t))$ is 1 when $\underline{w}_{ijk\ell}(x(t),y_{\xi}(t))$ and $\bar{w}_{ijk\ell}(x(t),y_{\xi}(t))$
are within the sub-FOU $\ell$; otherwise $\vartheta_{ijk\ell}(x(t),y_{\xi}(t))$ is 0.
In addition,
$\underline{\beta}_{ijk\ell}(x(t),y_{\xi}(t))$ and $\bar{\beta}_{ijk\ell}(x(t),y_{\xi}(t))$ are not necessarily known
but satisfy $\underline{\beta}_{ijk\ell}(x(t),y_{\xi}(t)) + \bar{\beta}_{ijk\ell}(x(t),y_{\xi}(t)) = 1$
and $0 \leq \underline{\beta}_{ijk\ell}(x(t),y_{\xi}(t))$, $\bar{\beta}_{ijk\ell}(x(t),y_{\xi}(t)) \leq 1$.

Later in this article, denote $\vartheta_{ijk\ell}(x(t),y_{\xi}(t))$,
$\underline{\beta}_{ijk\ell}(x(t),y_{\xi}(t))$,
$\underline{w}_{ijk\ell}(x(t),y_{\xi}(t))$,
$\bar{\beta}_{ijk\ell}(x(t),y_{\xi}(t))$,
$\bar{w}_{ijk\ell}(x(t),y_{\xi}(t))$,
$v_{\imath\jmath i_{\imath} j_{\jmath} ls\ell}\left(x_{\imath}(t),{y_{\xi}}_{\jmath}(t)\right)$
as
$\vartheta_{ijk\ell}$,
$\underline{\beta}_{ijk\ell}$,
$\underline{w}_{ijk\ell}$,
$\bar{\beta}_{ijk\ell}$,
$\bar{w}_{ijk\ell}$,
$v_{\imath\jmath i_{\imath} j_{\jmath} ls\ell}$, and
$\underline{o}_{ijk i_{1} i_{2} \cdots i_{n_{x}} j_{1} j_{2} \cdots j_{n_{y}} ls\ell}$,
$\bar{o}_{ijk i_{1} i_{2} \cdots i_{n_{x}} j_{1} j_{2} \cdots j_{n_{y}} ls\ell}$
as
$\underline{o}$,
$\bar{o}$
for simplicity, respectively.

Taking the nonnegative property of membership functions into consideration and introducing the slack matrices $W$ and $M_{ls\ell}$, one has
%utilizing the properties of membership functions, and considering about the slack matrices $W$ and $M_{ls\ell}$, the following inequality holds
\begin{equation}\begin{aligned}
\left\{\sum_{i \in \mathscr{I}} \sum_{j \in \mathscr{J}} \sum_{k \in \mathscr{I}} \sum_{\ell=1}^{\wp}
    \vartheta_{ijk\ell} \left( \underline{\beta}_{ijk\ell} \underline{w}_{ijk\ell} + \bar{\beta}_{ijk\ell} \bar{w}_{ijk\ell} \right) \right. \\ \left. -1\right.{\Bigg\}} W=0,
\end{aligned}     \label{MainT21}\end{equation}
\begin{equation}\begin{aligned}
\sum_{i \in \mathscr{I}} \sum_{j \in \mathscr{J}} \sum_{k \in \mathscr{I}} \sum_{\ell=1}^{\wp}
    \vartheta_{ijk\ell} \left( 1-\bar{\beta}_{ijk\ell} \right) \left( \bar{w}_{ijk\ell}-\underline{w}_{ijk\ell} \right) \\ \times M_{ijk\ell} \geq 0.
\end{aligned}     \label{MainT22}\end{equation}

Adding (\ref{MainT21}) and (\ref{MainT22}) into the LHS of (\ref{Main4}), the following inequality holds,
$$\begin{aligned}
%$$\begin{aligned}
\sum_{i \in \mathscr{I}}& \sum_{j \in \mathscr{J}} \sum_{k \in \mathscr{I}} m_{i}(x(t)) n_{j}(y_{\xi}(t)) m_{k}(x(t))
\Psi_{ijk}\\
=& \sum_{i \in \mathscr{I}} \sum_{j \in \mathscr{J}} \sum_{k \in \mathscr{I}} \sum_{\ell=1}^{\wp}
    \vartheta_{ijk\ell} \left( \underline{\beta}_{ijk\ell} \underline{w}_{ijk\ell} + \bar{\beta}_{ijk\ell} \bar{w}_{ijk\ell} \right)
    \Psi_{ijk}\\
%\end{aligned}$$
%\qquad\qquad\qquad\qquad\qquad\qquad\quad
\end{aligned}$$
\begin{equation}\begin{aligned}
\leq &\sum_{i \in \mathscr{I}} \sum_{j \in \mathscr{J}} \sum_{k \in \mathscr{I}} \sum_{\ell=1}^{\wp}
    \vartheta_{ijk\ell}
    \left( \left( 1-\bar{\beta}_{ijk\ell}\right) \underline{w}_{ijk\ell}\right.\\&\left.
    + \bar{\beta}_{ijk\ell} \bar{w}_{ijk\ell} \right)
    \left( \Psi_{ijk} + W \right) - W \\
& + \sum_{i \in \mathscr{I}} \sum_{j \in \mathscr{J}} \sum_{k \in \mathscr{I}} \sum_{\ell=1}^{\wp}
    \vartheta_{ijk\ell}
    \left( 1-\bar{\beta}_{ijk\ell} \right)\\&
    \left( \bar{w}_{ijk\ell} - \underline{w}_{ijk\ell} \right)
    M_{ijk\ell} \\
=&  \sum_{i \in \mathscr{I}} \sum_{j \in \mathscr{J}} \sum_{k \in \mathscr{I}} \sum_{\ell=1}^{\wp}
    \vartheta_{ijk\ell}
    \left( \underline{w}_{ijk\ell} \Psi_{ijk} + \left( \bar{w}_{ijk\ell} \right. \right. \\& \left.\left. - \underline{w}_{ijk\ell} \right) M_{ijk\ell}
    + \underline{w}_{ijk\ell} W \right) - W \\
& + \sum_{i \in \mathscr{I}} \sum_{j \in \mathscr{J}} \sum_{k \in \mathscr{I}} \sum_{\ell=1}^{\wp}
    \vartheta_{ijk\ell} \bar{\beta}_{ijk\ell}
    \left( \bar{w}_{ijk\ell} - \underline{w}_{ijk\ell} \right)\\
    &\times \left( \Psi_{ijk} - M_{ijk\ell} + W \right).
\end{aligned}     \label{MainT23}\end{equation}

It follows from (\ref{MainT23}) that the inequality in (\ref{Main4}) can be guaranteed by
\begin{equation}\begin{aligned}\label{MainT24}
    \sum_{i \in \mathscr{I}} \sum_{j \in \mathscr{J}} \sum_{k \in \mathscr{I}}
    \left( \underline{w}_{ijk\ell} \Psi_{ijk} + \left( \bar{w}_{ijk\ell} - \underline{w}_{ijk\ell} \right) M_{ijk\ell} \right. \\  \left.
    + \underline{w}_{ijk\ell} W \right) - W <0,
\end{aligned}\end{equation}
\begin{equation}
    \Psi_{ijk} - M_{ijk\ell} + W  < 0 .
\end{equation}

Substituting $\bar{w}_{ijk\ell}, \underline{w}_{ijk\ell}$ of (\ref{q26}) into (\ref{MainT24}),
which can be rewritten as

\begin{equation}\begin{aligned}
\sum_{l=1}^{p} \sum_{s=1}^{q}\sum_{i_{1}=1}^{2} \cdots \sum_{i_{n_{x}}=1}^{2} \sum_{j_{1}=1}^{2} \cdots \sum_{j_{n_{y}}=1}^{2}  \prod_{\imath=1}^{n_{x}} \prod_{\jmath=1}^{n_{y}}
    v_{\imath\jmath i_{\imath} j_{\jmath} ls\ell} \\
 \times\left(\sum_{i \in \mathscr{I}} \sum_{j \in \mathscr{J}} \sum_{k \in \mathscr{I}}
    \left( \underline{o} \Psi_{ijk} + \left( \bar{o} - \underline{o} \right) M_{ijk\ell}
    + \underline{o} W \right)\right) \\ - W < 0. \label{q49}
\end{aligned}\end{equation}

Then, the inequality (\ref{q49}) can be deduced by (\ref{MainT25}).
%It follows from \emph{Theorems 2} and \emph{Theorems 3} that the claimed results can be obtained if the LMIs in (\ref{q18}) and (\ref{q27}).
The proof is thus completed.
\hfill $\square$%$\Box$

To obtain the proposed MFD MDETM MOF controller, the above theorems will be utilized into the following optimization problem.

\textbf{Algorithm 1.}
%min $\gamma$, subject to (\ref{MainT27}),  (\ref{MainT25}),  (\ref{MainT26}),  with $\gamma>0$, $\delta>0$.
$$
\begin{gathered}
\min \gamma, \\
\text { subject to }(\ref{MainT27}),  (\ref{MainT25}),  (\ref{MainT26}), \\
\text { with } \gamma>0, \quad \theta>0 .
\end{gathered}
$$

Similarly, the proposed MFI MDETM MOF controller can be obtained by the following convex optimization algorithm.

\textbf{Algorithm 2.}
%min $\gamma$, subject to (\ref{Main0}), (\ref{Main1}),  with $\gamma>0$, $\delta>0$.
$$
\begin{gathered}
\min \gamma, \\
\text { subject to }(\ref{Main0}), (\ref{Main1}), \\
\text { with } \gamma>0, \theta>0 .
\end{gathered}
$$

\textbf{Remark 2.}
The MDETM MOF controller design results are given in Algorithm 1 or Algorithm 2 for any positive integer $\kappa\in N$.
As a special case, a traditional DETM memoryless controller design method can be obtained by setting $\kappa=1$ in Algorithm 1 or Algorithm 2.
Similarly, by setting $\varrho_{1}=1$ and $\varrho_{h}=0$ $\left(for \ h=2,...,\kappa\right)$ in Algorithm 1 or Algorithm 2, the MDETM MOF controller design approach reduces to a DETM MOF controller design method.

%\textbf{Remark 2:}
%The simulation below shows that the utilization of additional past system output measurements can improve the $\mathscr H_{\infty}$ performance, which becomes better as the number of past data increases. However, meanwhile, the computational costs also increase with the increasing number of past system information used for the controller design. In practice, the balance between design conservatism and computational complexity should be considered carefully.

\textbf{Remark 3.}
Simulation results indicate that the inclusion of additional past system output measurements significantly improves $\mathscr{H}_{\infty}$ performance, with enhancements becoming more substantial as the number of past data points increases. However, this improvement is accompanied by an increase in computational complexity. Consequently, in practical applications, careful consideration must be given to balancing reduced design conservatism with manageable computational requirements.

\section{Simulation Studies}
\textbf{Example 1.} \label{example 1}
Consider a numerical example with 3-rules in the form of (\ref{q1}), which can be described below: %\textcolor{blue}{
\begin{equation*}
\begin{aligned}
\begin{array}{lllllllll}
A_{1}=A_{3}=\left[\begin{array}{cc}1.1 &0\\-0.3&0.3\end{array}\right], \\
A_{2}=\left[\begin{array}{cc}0.86 &0\\-0.2&0.1\end{array}\right],\\
B_{u1}=B_{u3}=\left[\begin{array}{cc}1.1\\0.3\end{array}\right],\\
B_{u2}=\left[\begin{array}{cc}1.2\\0.6\end{array}\right],\\
B_{d1}=B_{d2}=B_{d3}=\left[\begin{array}{cc}1\\1\end{array}\right],\\
C_{y1}=C_{y2}=C_{y3}=\left[\begin{array}{cc}1&0\end{array}\right],\\
C_{z1}=C_{z2}=C_{z3}=\left[\begin{array}{cc}0.05&0\end{array}\right]\\
\end{array}
\end{aligned}
\end{equation*}

where $x_1(t)\in[-4,4]$, $y(t)\in[-4,4]$, and the membership functions are defined as follows:
%$\underline{w}_{1}\left(x_{1}\right)=\left(\left[x_{1}(t)+3\right] / 10\right)$,
%$\bar{w}_{1}\left(x_{1}\right)=\left(\left[x_{1}(t)+4\right] / 10\right)$,
%$\underline{w}_{2}\left(x_{1}\right) = 0.4$,
%$\bar{w}_{2}\left(x_{1}\right) = 0.5$,
%$\underline{w}_{3}\left(x_{1}\right) = \left(\left[-x_{1}(t)+3\right] / 10\right)$,
%$\bar{w}_{3}\left(x_{1}\right) = \left(\left[-x_{1}(t)+5\right] / 10\right)$,
\begin{align*}
\begin{array}{lllllllll}
\underline{w}_{1}\left(x_{1}\right)=\left(\left[x_{1}(t)+3\right] / 10\right), \\
\bar{w}_{1}\left(x_{1}\right)=\left(\left[x_{1}(t)+4\right] / 10\right), \\
\underline{w}_{2}\left(x_{1}\right) = 0.4,
\bar{w}_{2}\left(x_{1}\right) = 0.5, \\
\underline{w}_{3}\left(x_{1}\right) = \left(\left[-x_{1}(t)+3\right] / 10\right), \\
\bar{w}_{3}\left(x_{1}\right) = \left(\left[-x_{1}(t)+5\right] / 10\right),
\end{array}
\end{align*}
and the final membership functions are given by:
\begin{align*}
\begin{array}{lllllllll}
w_{1}(x_{1}) =& (1 - \sin^2(0.4 x_{1}(t))) \underline{w}_{1}(x_{1}) \\ &+ \sin^2(0.4 x_{1}(t)) \bar{w}_{1}(x_{1}), \\
w_{2}(x_{1}) =& 0.5 \underline{w}_{2}(x_{1}) + 0.5 \bar{w}_{2}(x_{1}), \\
w_{3}(x_{1}) =& \sin^2(0.4 x_{1}(t)) \underline{w}_{3}(x_{1}) \\ &+ (1 - \sin^2(0.4 x_{1}(t))) \bar{w}_{3}(x_{1}).
\end{array}
\end{align*}
%and
%$w_{1}\left(x_{1}\right)=
%    \left(1-\sin ^{2} \left(0.4x_{1}(t)\right)\right)\underline{w}_{1}\left(x_{1}\right)
%    +\sin ^{2} \left(0.4x_{1}(t)\right)\bar{w}_{1}\left(x_{1}\right)$,
%$w_{2}\left(x_{1}\right)=
%    0.5 \underline{w}_{2}\left(x_{1}\right)
%    +0.5\bar{w}_{2}\left(x_{1}\right)$,
%$w_{3}\left(x_{1}\right)=
%    \left(\sin ^{2} \left(0.4x_{1}(t)\right)\right)
%    \underline{w}_{3}\left(x_{1}\right)
%    +\left(1-\sin ^{2} \left(0.4x_{1}(t)\right)\right)
%    \bar{w}_{3}\left(x_{1}\right)$.
In this study, the stochastic variable $\xi$ is characterized by its mean value $\bar{\xi} = 0.8$ and standard deviation $\xi^* = 0.05$. The failure parameter is set to $\alpha_f = 1$, and the disturbance input is given by $d(t) = 3 e^{-0.1t} \sin(t)$.

\begin{table*}
    \centering
    %TABLE \MakeUppercase{\romannumeral 1}\\
    %\caption{THE $H_{\infty}$ PERFORMANCE $\gamma_{\min}$ OBTAINED BY DIFFERENT CASES}
    \caption{The $\mathscr H_{\infty}$ Performance $\gamma_{\min}$ Obtained by Different Cases}
%    THE LOWER AND UPPER MEMBERSHIP FUNCTIONS\\
%    The $H_\infty$ performance $\gamma_{\min}$ obtained by different Cases\\
    \begin{tabular}{c|c|c|c|c|c|c|c|c|c|c}
        \hline \hline
         $\kappa$   & 1 & 2 & 3 & 4 & 5 & 6 & 7 & 8 & 9 & 10
            \\%1.2 48.3618      0.30119421191220209664497760708322
        \hline
           Case 1 & 23.493 & - & - & - & - & - & - & - & - & -
            \\%1.2 48.3618      0.30119421191220209664497760708322
        \hline
           Case 2 & 23.493 & 9.073 & 8.836 & 8.580 & 8.357 & 8.136 & 7.930 & 7.739 & 7.549 & 7.377
            \\%1.1              0.33287108369807955328884690643132
        \hline
           Case 3 & 23.493 & 8.637 & 8.361 & 8.142 & 7.942 & 7.725 & 7.528 & 7.354 & 7.149 & 6.989
            \\%1.1              0.33287108369807955328884690643132
        \hline
           Case 4 & 0.795 & 0.526 & 0.525 & 0.524 & 0.523 & 0.522 & 0.522 & 0.521 & 0.521 & 0.521
            \\%1.1              0.33287108369807955328884690643132
        \hline \hline
    \end{tabular}
\end{table*}

\begin{table*}
    \centering
    %TABLE \MakeUppercase{\romannumeral 2}\\
    %THE $H_{\infty}$ PERFORMANCE $\gamma_{\min}$ OBTAINED BY DIFFERENT CASES\\
    %THE LOWER AND UPPER MEMBERSHIP FUNCTIONS\\
    %\caption{THE TRIGGERING RATES BY DIFFERENT CASES}
    \caption{The Triggering Rates by Different Cases}
    \begin{tabular}{c|c|c|c|c|c|c|c|c|c|c}
        \hline \hline
         $\kappa$   & 1 & 2 & 3 & 4 & 5 & 6 & 7 & 8 & 9 & 10
            \\%1.2 48.3618      0.30119421191220209664497760708322
        \hline
           Case 1 & 0.23 & - & - & - & - & - & - & - & - & -
            \\%1.2 48.3618      0.30119421191220209664497760708322
        \hline
           Case 2 & 0.23 & 0.23 & 0.23 & 0.28 & 0.26 & 0.28 & 0.28 & 0.28 & 0.38 & 0.37
            \\%1.1              0.33287108369807955328884690643132
        \hline
           Case 3 & 0.23 & 0.22 & 0.22 & 0.20 & 0.19 & 0.19 & 0.18 & 0.17 & 0.17 & 0.17
            \\%1.1              0.33287108369807955328884690643132
        \hline
           %Case 4 & 0.21 & 0.23 & 0.22 & 0.20 & 0.17 & 0.15 & 0.19 & 0.19 & - & -
           Case 4 & 0.23 & 0.21 & 0.22 & 0.20 & 0.19 & 0.17 & 0.17 & 0.19 & 0.15 & 0.15
            \\%1.1              0.33287108369807955328884690643132
        \hline \hline
    \end{tabular}
\end{table*}

The objective is to design a three-rule interval type-2 (IT-2) fuzzy controller in the form of (\ref{Pre7}) such that the closed-loop system achieves mean-square exponential stability with guaranteed $\mathscr{H}_{\infty}$ performance.

%%The membership functions of controller are set as
%First, the upper and lower bound membership functions for the controller are defined as follows:
%$ \underline{m}_{1}\left(y_{\xi}\right)=\left(\left[y_{\xi}(t)+3\right] / 10\right), \quad \bar{m}_{1}\left(y_{\xi}\right)=\left(\left[y_{\xi}(t)+4\right] / 10\right), $
%$ \underline{m}_{2}\left(y_{\xi}\right)=\left(\left[-y_{\xi}(t)+3\right] / 10\right), \quad \bar{m}_{2}\left(y_{\xi}\right)=\left(\left[-y_{\xi}(t)+5\right] / 10\right). $
%The precise membership functions are then obtained based on the upper and lower bound membership functions through fixed parameters:
%$ m_{1}\left(y_{\xi}\right)=
%    \left(1-\sin ^{2} \left(0.4y_{\xi}(t)\right)\right)\underline{m}_{1}\left(y_{\xi}\right)
%    +\sin ^{2} \left(0.4y_{\xi}(t)\right) \bar{m}_{1}\left(y_{\xi}\right), $
%$ m_{2}\left(y_{\xi}\right)=
%    0.5 \underline{m}_{2}\left(y_{\xi}\right)
%    +0.5\bar{m}_{2}\left(y_{\xi}\right), $
%$ m_{3}\left(y_{\xi}\right)=
%    1-m_{1}\left(y_{\xi}\right)-m_{2}\left(y_{\xi}\right). $

First, the upper and lower bound membership functions for the controller are defined as follows:
\begin{align*}
\begin{array}{l}
\underline{m}_{1}\left(y_{\xi}\right)=\left(\left[y_{\xi}(t)+3\right] / 10\right), \\
\bar{m}_{1}\left(y_{\xi}\right)=\left(\left[y_{\xi}(t)+4\right] / 10\right), \\
\underline{m}_{2}\left(y_{\xi}\right)=\left(\left[-y_{\xi}(t)+3\right] / 10\right), \\
\bar{m}_{2}\left(y_{\xi}\right)=\left(\left[-y_{\xi}(t)+5\right] / 10\right).
\end{array}
\end{align*}

Based on these bounds, the precise membership functions are obtained through fixed parameters as follows:
\begin{align*}
m_{1}(y_{\xi}) &= (1 - \sin^2(0.4 y_{\xi}(t))) \underline{m}_{1}(y_{\xi}) \\
               &\quad + \sin^2(0.4 y_{\xi}(t)) \bar{m}_{1}(y_{\xi}), \\
m_{2}(y_{\xi}) &= 0.5 \underline{m}_{2}(y_{\xi})
                 + 0.5 \bar{m}_{2}(y_{\xi}), \\
m_{3}(y_{\xi}) &= 1 - m_{1}(y_{\xi}) - m_{2}(y_{\xi}).
\end{align*}

%$\underline{m}_{1}\left(y_{\xi}\right)=\left(\left[y_{\xi}(t)+3\right] / 10\right)$,
%$\bar{m}_{1}\left(y_{\xi}\right)=\left(\left[y_{\xi}(t)+4\right] / 10\right)$,
%$\underline{m}_{2}\left(y_{\xi}\right)=\left(\left[-y_{\xi}(t)+3\right] / 10\right)$,
%$\bar{m}_{2}\left(y_{\xi}\right)=\left(\left[-y_{\xi}(t)+5\right] / 10\right)$,
%$m_{1}\left(y_{\xi}\right)=
%    \left(1-\sin ^{2} \left(0.4y_{\xi}(t)\right)\right)\underline{m}_{1}\left(y_{\xi}\right)
%    +\sin ^{2} \left(0.4y_{\xi}(t)\right) \bar{m}_{1}\left(y_{\xi}\right)$,
%$m_{2}\left(y_{\xi}\right)=
%    0.5 \underline{m}_{2}\left(y_{\xi}\right)
%    +0.5\bar{m}_{2}\left(y_{\xi}\right)$,
%$m_{3}\left(y_{\xi}\right)=
%    1-m_{1}\left(y_{\xi}\right)-m_{2}\left(y_{\xi}\right)$.

Considering the membership function dependency method in Theorem 3.2, the premise variable's domain is divided into $ m = n = 20 $ subspaces, $\mathcal{S} = \cup_{\imath=1}^{m} \cup_{\jmath=1}^{n} \mathcal{S}_{\imath\jmath}$, where $\imath \in \{1, 2, \cdots, m\}$ and $\jmath \in \{1, 2, \cdots, n\}$, with $ m = n = 20 $. Each subspace $\mathcal{S}_{\imath\jmath}$ is specifically divided as follows:
$$
\underline{x}_{1_\imath} = -4 + \frac{8(\imath-1)}{m}, \quad \overline{x}_{1_\imath} = -4 + \frac{8\imath}{m},
$$
$$
\underline{y}_{\xi_\jmath} = -4 + \frac{8(\jmath-1)}{n}, \quad \overline{y}_{\xi_\jmath} = -4 + \frac{8\jmath}{n}.
$$

According to Theorem 2, only one FOU is divided, i.e., $\wp  = 1$, and the other fixed parameters can be set as follows:
$$
\begin{array}{ccccc}
{\underline{\varepsilon}_{ls11\imath \jmath }}   = {\underline{w}_l}\left( {{{\underline {{x_1}} }_\imath }} \right){\underline{m}_s}({\underline {{y_\xi }} _\jmath }),\quad
{\underline{\varepsilon}_{ls12\imath \jmath }} = {\underline{w}_l}\left( {{{\underline {{x_1}} }_\imath }} \right){\underline{m}_s}({\underline {{y_\xi }} _\jmath }),\\
{\underline{\varepsilon}_{ls21\imath \jmath }}   = {\underline{w}_l}\left( {{{\overline {{x_1}} }_\imath }} \right){\underline{m}_s}({\overline {{y_\xi }} _\jmath }),\quad
{\underline{\varepsilon}_{ls22\imath \jmath }} = {\underline{w}_l}\left( {{{\overline {{x_1}} }_\imath }} \right){\underline{m}_s}({\overline {{y_\xi }} _\jmath }),\\
{\bar{\varepsilon}_{ls11\imath \jmath }}   = {\bar{w}_l}\left( {{{\underline {{x_1}} }_\imath }} \right){\bar{m}_s}({\underline {{y_\xi }} _\jmath }),\quad
{\bar{\varepsilon}_{ls12\imath \jmath }} = {\bar{w}_l}\left( {{{\underline {{x_1}} }_\imath }} \right){\bar{m}_s}({\underline {{y_\xi }} _\jmath }),\\
{{\bar \varepsilon }_{ls21\imath \jmath }}   = {{\bar w}_l}\left( {{{\overline {{x_1}} }_\imath }} \right){{\bar m}_s}({\overline {{y_\xi }} _\jmath }),\quad
{{\bar \varepsilon }_{ls22\imath \jmath }} = {{\bar w}_l}\left( {{{\overline {{x_1}} }_\imath }} \right){{\bar m}_s}({\overline {{y_\xi }} _\jmath }),\\
l \in {\cal L}  ,\quad s \in {\cal S},\quad \imath  \in \{ 1,2, \cdots ,m\} ,\quad \jmath  \in \{ 1,2, \cdots ,n\} .
\end{array}
$$

%where $ l \in \mathcal{L} $, $ s \in \mathcal{S} $, $ \imath \in \{1, 2, \cdots, m\} $, and $ \jmath \in \{1, 2, \cdots, n\} $.

The following section will compare the controllers obtained through Algorithm 1 or Algorithm 2 for different scenarios in terms of $\mathscr H_\infty$ performance, convergence speed, and the number of event triggers:

%By applying Algorithm~1 and Algorithm~2, feasible solutions can be obtained, leading to optimal performance metrics. Table~\MakeUppercase{\romannumeral 1} provides a detailed comparison of the performance for various control strategies:

$$
\left\{
\begin{array}{ll}
\text{Case 1:} & \text{MFI DETM OF controller}; \\
\text{Case 2:} & \text{MFI DETM MOF controller}; \\
\text{Case 3:} & \text{MFI MDETM MOF controller}; \\
\text{Case 4:} & \text{MFD MDETM MOF controller}
\end{array}
\right.
$$

%\textbf{Abbreviations:}
where
\begin{itemize}
    \item MFI: Membership Function Independent
    \item MFD: Membership Function Dependent
    \item DETM: Dynamic Event-Triggered Mechanism, Memoryless
    \item MDETM: Memory Dynamic Event-Triggered Mechanism
    \item OF: Output-Feedback, Memoryless
    \item MOF: Memory Output Feedback
\end{itemize}

In Table \MakeUppercase{\romannumeral 1}, comparing Case 1 and Case 2, it can be observed that as the number of stored historical time points $\kappa$ increases, the controller's $\mathscr H_\infty$ minimum performance index $\gamma_{\min}$ can become smaller. This indicates that compared to traditional memoryless $\mathscr H_\infty$ robust controllers, memory-based $\mathscr H_\infty$ robust controllers have enhanced $\mathscr H_\infty$ performance. Comparing Case 2 and Case 3, it can be observed that memory-based dynamic event-triggering mechanisms and ordinary dynamic event-triggering mechanisms have similar $\mathscr H_\infty$ performance, with memory-based dynamic event-triggering mechanisms achieving slightly smaller minimum $\mathscr H_\infty$ performance indices $\gamma_{\min}$. Comparing Case 3 and Case 4, it can be observed that the piecewise membership function dependency technique adopted in this chapter can significantly reduce the $\mathscr H_\infty$ performance index $\gamma_{\min}$, thereby enhancing the system's $\mathscr H_\infty$ performance.

\begin{table}[h!]
\centering
\caption{Parameter Values for Different Cases}
\label{tab:parameters}
\renewcommand{\arraystretch}{1.5} % Increase row height for readability
\begin{tabular}{@{}l *{3}{>{$}c<{$}} @{}}
\toprule
\textbf{Case} & \tilde{\varrho} & E & F \\
\midrule
1 & I_{n_x} & 0.17I_{n_x} & I_{n_x} \\
2 & \left[\begin{smallmatrix} 0_{n_x} & 0_{n_x} & 0_{n_x} & I_{n_x} \end{smallmatrix}\right]
  & \left[\begin{smallmatrix} 0.01 \\ 0.01 \\ 0.01 \\ 0.17 \end{smallmatrix}\right] I_{n_x}
  & I_{n_x} \\
3 & \left[\begin{smallmatrix} 0.01 & 0.07 & 0.08 & 0.84 \end{smallmatrix}\right] I_{n_x}
  & \left[\begin{smallmatrix} 0.01 \\ 0.01 \\ 0.15 \\ 0.3 \end{smallmatrix}\right] I_{n_x}
  & I_{n_x} \\
4 & \left[\begin{smallmatrix} 0.01 & 0.07 & 0.08 & 0.84 \end{smallmatrix}\right] I_{n_x}
  & \left[\begin{smallmatrix} 0.01 \\ 0.01 \\ 0.15 \\ 0.3 \end{smallmatrix}\right] I_{n_x}
  & I_{n_x} \\
\bottomrule
\end{tabular}
\end{table}

\begin{figure*}[!ht]
    \centering
    \includegraphics[width=0.7\textwidth]{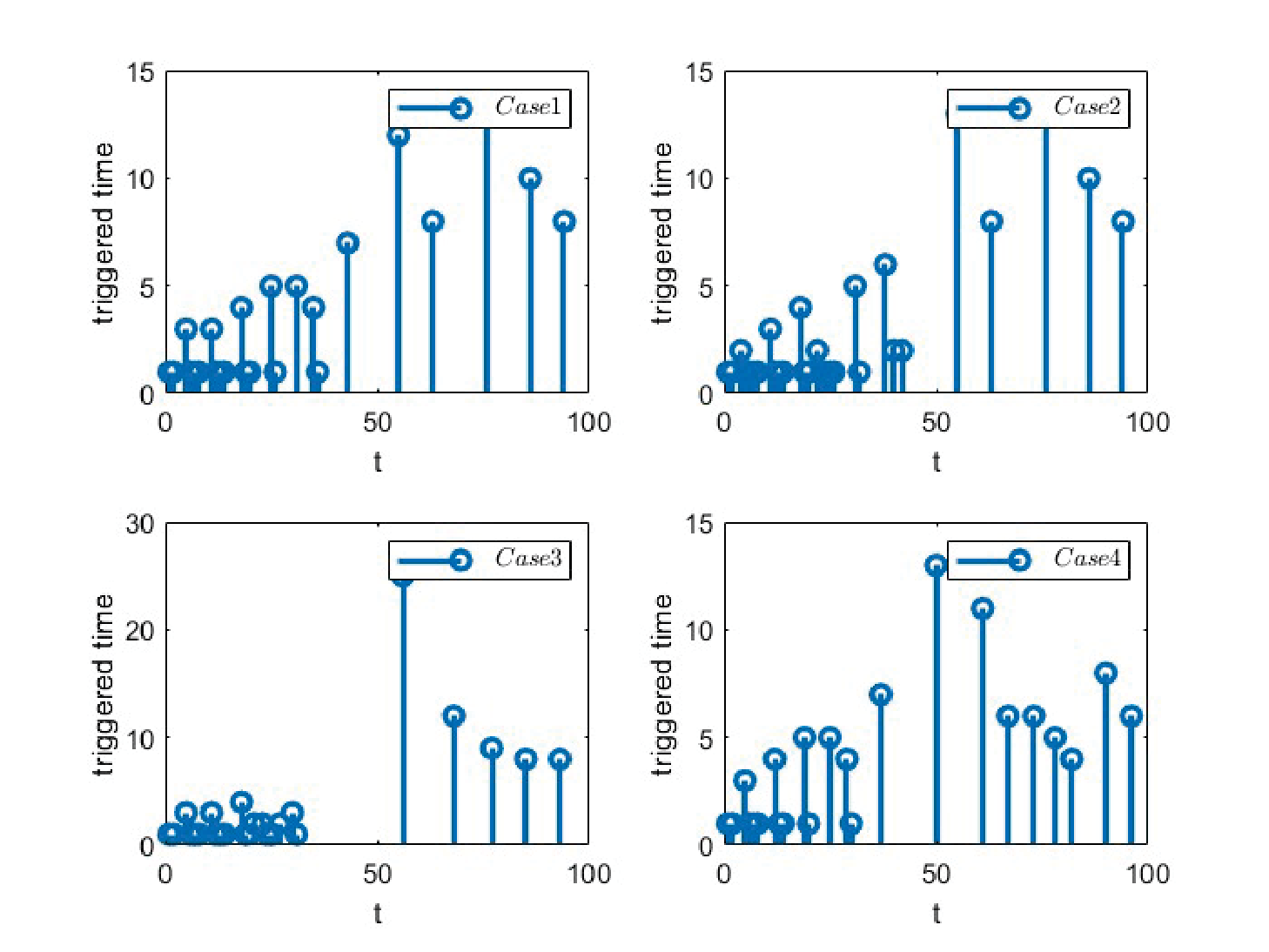}
    \caption{ Triggered time of different Cases} % 添加描述性的标题
    \label{fig:network_structure_3_2} % 确保标签唯一
\end{figure*}

In Table \MakeUppercase{\romannumeral 2}, the triggering rates (TRs) are a metric used to measure the frequency of event triggers, calculated as follows:

$$
\text{TRs} = \frac{\text{Total number of event trigger times}}{\text{Total number of sampling times}}
$$

Comparing Case 1 and Case 2, it can be observed that as the number of stored historical time points $\kappa$ increases, the system's TRs also increase. This indicates that under the same dynamic event-triggering mechanism, a memory-based controller leads to more event triggers compared to a traditional memoryless controller. Comparing Case 2 and Case 3, it can be observed that, compared to ordinary dynamic event-triggering mechanisms, the TRs under a memory-based dynamic event-triggering mechanism generally decrease as $\kappa$ increases. Comparing Case 3 and Case 4, it can be observed that the piecewise membership function dependency technique adopted in this chapter can also reduce the system's TRs to some extent.

\begin{figure}[!ht]
    \centering
    \includegraphics[width=0.45\textwidth]{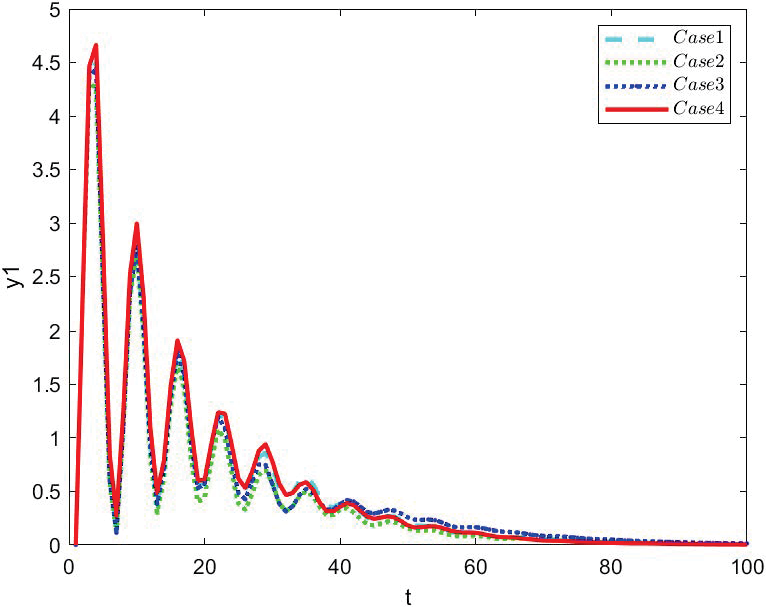}
    \caption{ Output signal ${y}_{1}$ responses of different Cases} % 添加描述性的标题
    \label{fig:network_structure_3_3} % 确保标签唯一
\end{figure}

%More specifically, by Algorithm 1 (h=4), one can obtain feasible solutions with $\alpha=0.01$, $\nu$, $\rho$, $\mu$, $F$, and
%$$
%\begin{aligned}
%\left\{
%\begin{array}{lll}
%Case 1: &E  &\tilde{\varrho};\\
%Case 2: & & ;\\
%Case 3: & & ;\\
%Case 4: & & ;\\
%Case 5: & & ;\\
%Case 6: & & .
%\end{array}
%\right.
%\end{aligned}
%$$
%and the controller gains as
%$$
%\begin{aligned}
%\left\{
%\begin{array}{ll}
%Case 1: &\Omega;\\
%Case 2: &;\\
%Case 3: &;\\
%Case 4: &;\\
%Case 5: &;\\
%Case 6: &.
%\end{array}
%\right.
%\end{aligned}
%$$

To observe the detailed event triggering conditions of the system as well as the response curves of the output signals and control signals, specifically, for Case 2, Case 3, and Case 4, let $\kappa = 4$, and some other parameter values are shown in Table~\ref{tab:parameters}.

\begin{figure}[!ht]
    \centering
    \includegraphics[width=0.45\textwidth]{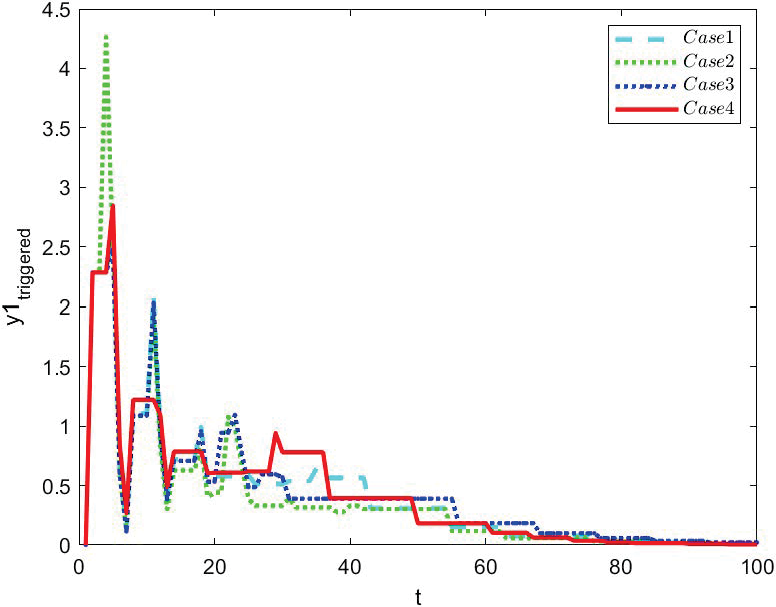}
    \caption{ Triggered output signal ${{\rm{y}}_1}_{triggered}$ responses of different Cases} % 添加描述性的标题
    \label{fig:network_structure_3_4} % 确保标签唯一
\end{figure}

\begin{figure}[!ht]
    \centering
    \includegraphics[width=0.45\textwidth]{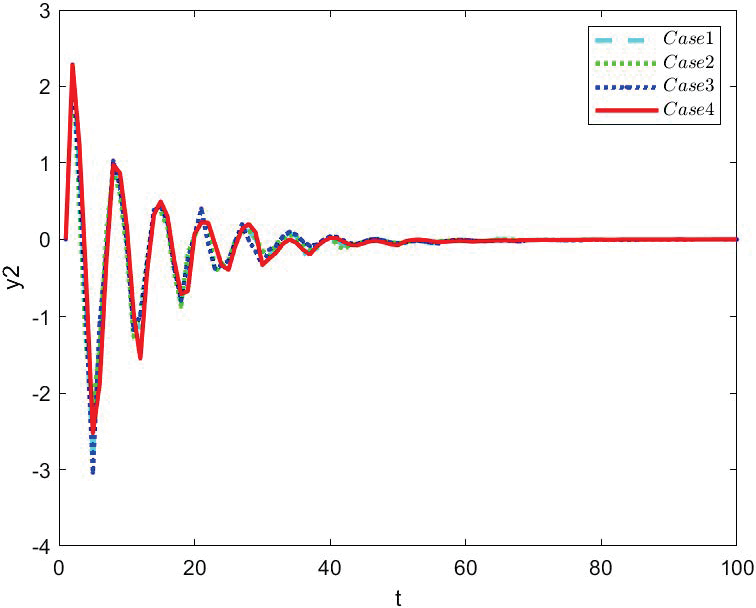}
    \caption{ Output signal ${y}_{1}$ responses of different Cases} % 添加描述性的标题
    \label{fig:network_structure_3_5} % 确保标签唯一
\end{figure}

In the specific cases shown in Table~\ref{tab:parameters}, the minimum performance indices $\gamma_{\min}$ for the $\mathscr H_\infty$ performance of Case 1, Case 2, Case 3, and Case 4 are 23.4932, 8.5799, 4.8512, and 0.5194, respectively. This result aligns with the conclusions drawn in the previous section: compared to traditional memoryless $\mathscr H_\infty$ output feedback controllers, memory-based $\mathscr H_\infty$ output feedback controllers exhibit stronger robustness; under a memory-based dynamic event-triggering mechanism, the robustness of memory-based $\mathscr H_\infty$ output feedback controllers is slightly stronger than that under an ordinary dynamic event-triggering mechanism; and compared to memory-based dynamic event-triggering with independent membership functions, the memory-based dynamic event-triggering with piecewise membership function dependency technique results in even stronger robustness.

%The detailed event triggering conditions of the system as well as the response curves of the output signals and control signals are shown in Figure~\ref{fig:network_structure_3_2}.

Figure~\ref{fig:network_structure_3_2} shows the number of event triggers under different strategies. In Figure~\ref{fig:network_structure_3_2}, the TRs for Case 1, Case 2, Case 3, and Case 4 are 0.23, 0.28, 0.22, and 0.21, respectively. This generally aligns with the previously discussed results: due to the introduction of the memory-based output feedback control method, the TRs for Case 2 are higher than those for Case 1; then, due to the introduction of the memory-based dynamic event-triggering mechanism, the TRs for Case 3 are significantly lower than those for Case 2 and also lower than those for Case 1; finally, the piecewise membership function dependency technique further reduces the TRs for Case 4.

\begin{figure}[!ht]
    \centering
    \includegraphics[width=0.45\textwidth]{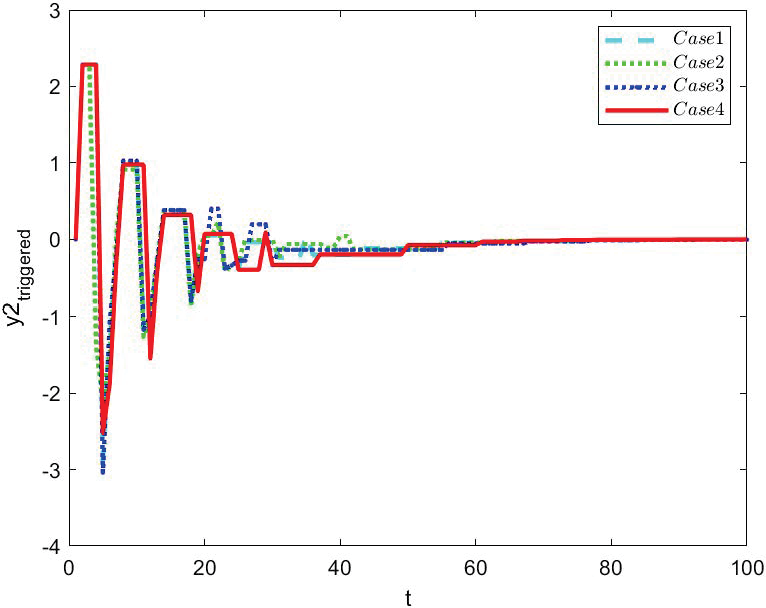}
    \caption{ Triggered output signal ${{\rm{y}}_2}_{triggered}$ responses of different Cases} % 添加描述性的标题
    \label{fig:network_structure_3_6} % 确保标签唯一
\end{figure}

\begin{figure}[!ht]
    \centering
    \includegraphics[width=0.45\textwidth]{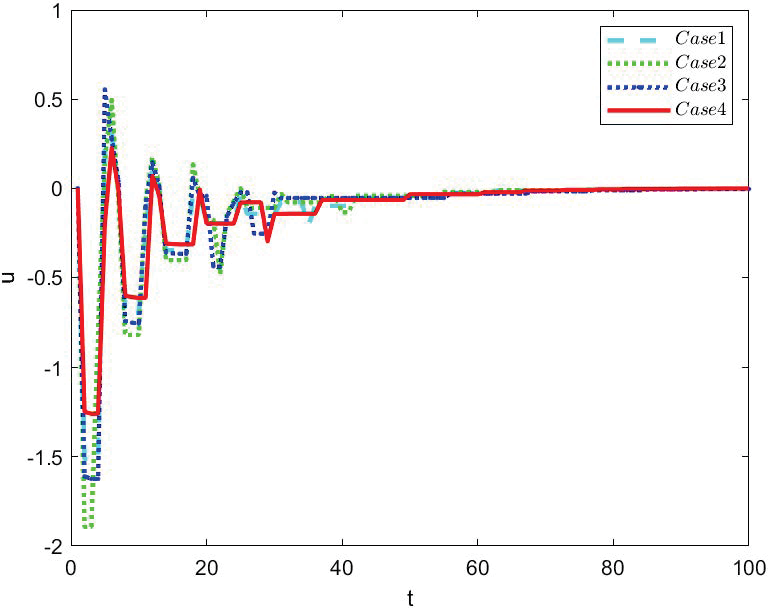}
    \caption{ Control signal responses of different Cases} % 添加描述性的标题
    \label{fig:network_structure_3_7} % 确保标签唯一
\end{figure}

Figure~\ref{fig:network_structure_3_3} and Figure~\ref{fig:network_structure_3_4} respectively describe the output signals $ \mathbf{y}_1 $ from the controlled system in Case 1, Case 2, Case 3, and Case 4, as well as the output signals $ \mathbf{y}_{1\text{triggered}} $ transmitted to the controller end after event triggering and random channel fading.

Figure~\ref{fig:network_structure_3_5} and Figure~\ref{fig:network_structure_3_6} respectively describe the output signals $ \mathbf{y}_2 $ from the controlled system in Case 1, Case 2, Case 3, and Case 4, as well as the output signals $ \mathbf{y}_{2\text{triggered}} $ transmitted to the controller end after event triggering and random channel fading.

Figure~\ref{fig:network_structure_3_7} describes the response curves of the control signals in Case 1, Case 2, Case 3, and Case 4.

By observing the response curves described in Figures~\ref{fig:network_structure_3_3}, \ref{fig:network_structure_3_4}, \ref{fig:network_structure_3_5}, \ref{fig:network_structure_3_6}, and~\ref{fig:network_structure_3_7}, it can be concluded that the different strategies have similar response curves and control performance, all of which are stable and effective. Among them, Case 4 has a slight advantage. Under the same performance, the strategy with a memory-based dynamic event-triggering mechanism and memory-based output feedback with membership function dependency can achieve superior $ \mathscr H_\infty $ performance with fewer trigger events.

\section{Conclusions}
%In this paper, the mean-square exponential stability and $\mathscr H_{\infty}$ performance $\gamma$ have been guaranteed for the proposed discrete-time MDETM MOF controller of  IT2 fuzzy systems with fading channel and actuator failure. During the controller design process, a MDETM is utilized to reduce the frequency of data transmission, and a MOF controller is utilized for reducing design conservation. Channel fading and actuator failure that may occur in the actual system are modeled and taken considering into the controller design conditions. Besides, the imperfectly matched the MFs and PDC strategy have been handled via an MFD stability method. From the simulations, it can be seen that the controller designed by the method generally improve the H∞ performance, reduce the triggered time and has a good resistance to channel fading. The advantages of our method over the memoryless and/or membership-function-independent (MFI) ones have been verified by simulation studies. In future works, some novel methods to further reduce the design conservatism would be proposed, and more systematic network control issues would be considered.
This paper presents a novel design method for a discrete-time MDETM MOF controller tailored for IT2 fuzzy systems. The approach ensures mean-square exponential stability and achieves satisfactory $\mathscr H_{\infty}$ performance $\gamma$ in the presence of channel fading and actuator failures. Leveraging MDETM minimizes data transmission frequency, while employing a MOF strategy reduces design conservatism. Real-world issues such as channel fading and actuator failures are addressed through comprehensive modeling and incorporated into the controller design criteria. The problem of imperfectly matched membership functions is handled using a Non-PDC mechanism and an MFD stability analysis method. Simulation results demonstrate that the designed controller improves robust performance, reduces triggering frequency, and exhibits strong resilience against channel fading. Compared to memoryless or membership-function-independent methods, this technique offers superior performance, as verified by extensive simulations. Future work will focus on further reducing design conservatism and addressing more complex networked control challenges.

\begin{appendix}

\textbf{Lemma A1.} \label{lemma 2} \textbf{\textup{\cite{S.Zhang2016}}} %\textbf{\textup{(\cite{S.Zhang2016} Lemma 5)}}
Suppose that there exist real matrices $\mathscr{M}$, $\mathscr{W}$, $\mathscr{X}$ and $\mathscr{Y}$, 
the following two statements are equivalent.
\begin{enumerate}[1)]
\item The following conditions hold:$$
    \left[\begin{array}{cc}
    \mathscr{M} & * \\
    \mathscr{W}^{T}+\epsilon \mathscr{Y} & -\epsilon \mathscr{X}-\epsilon \mathscr{X}^{T}
    \end{array}\right]<0,
    $$ where $\epsilon$ is a real constant scalar.l\\
\item The following inequality is solvable:$$
    \mathscr{M}+\mathscr{W} \mathscr{X}^{-1} \mathscr{Y}+\mathscr{Y}^{T} \mathscr{X}^{-T} \mathscr{W}^{T}<0.
    $$\\
\end{enumerate}

\textbf{Lemma A2.} \label{lemma 3} \textbf{\textup{\cite{S.Zhangb2018}}} Consider the given constant $0<\hbar<1$ and Lyapunov function $V(x(t))=x^{T}(t) P x(t)$, the discrete-time closed-loop fuzzy system (\ref{Pre7}) is exponentially stable in the mean-square sense, if the following inequality is satisfied:
%$$
%\mathbb{E}\{\Delta V(x(t))\} = \mathbb{E}\{V(x(t+1)) - V(x(t))\} \leq - \hbar \mathbb{E}\{V(x(t))\}.
%$$
\begin{align*}
\mathbb{E}\{\Delta V(x(t))\}
&= \mathbb{E}\{V(x(t+1)) - V(x(t))\} \\
&\leq - \hbar \mathbb{E}\{V(x(t))\}.
\end{align*}

Therefore, for $t \in [ t_l,t_{l+1} )$, it holds that
$$
\mathbb{E}\{V(x(t))\} \leq (1-\hbar)^{(t-t_l)} \mathbb{E}\{V(x(t_l))\}.
$$

Then,one can achieve that
$$
\begin{aligned}
\mathbb{E}\{V(x(t))\} &\leq (1-\hbar)^{(t-t_l)} \mathbb{E}\{V(x(t_l))\}\\
            &\leq \cdots\leq (1-\hbar)^{(t-t_0)} \mathbb{E}\{V(x(t_0))\}.
\end{aligned}
$$

On the other hand, for the Lyapunov function,
$$
a\mathbb{E}\{{\| x(t) \|}^{2}\} \leq \mathbb{E}\{V(x(t))\} \leq b\mathbb{E}\{{\| x(t) \|}^{2}\}
$$
with $a=\lambda_{min}(P)$ and $b=\lambda_{max}(P)$.

By the way,
%$$
%\mathbb{E}\{{\| x(t) \|}^{2}\} \leq
%\left(\frac{1}{a}\right)\mathbb{E}\{V(x(t))\} \leq
%\left(\frac{1}{a}\right) (1-\hbar)^{(t-t_0)} \mathbb{E}\{V(x(t_0))\} \leq
%\left(\frac{b}{a}\right) (1-\hbar)^{(t-t_0)} \mathbb{E}\{{\| x(t_0) \|}^{2}\}.
%$$

\begin{align*}
\mathbb{E}\{{\| x(t) \|}^{2}\}
&\leq \left(\frac{1}{a}\right)\mathbb{E}\{V(x(t))\} \nonumber \\
&\leq \left(\frac{1}{a}\right) (1-\hbar)^{(t-t_0)} \mathbb{E}\{V(x(t_0))\} \nonumber \\
&\leq \left(\frac{b}{a}\right) (1-\hbar)^{(t-t_0)} \mathbb{E}\{{\| x(t_0) \|}^{2}\}.
\end{align*}

Denoting $\hslash=(1-\hbar)$ and $c=\left(\frac{b}{a}\right)$,
$$
\mathbb{E}\{{\| x(t) \|}^{2}\} \leq
c \hslash^{(t-t_0)} \mathbb{E}\{{\| x(t_0) \|}^{2}\}.
$$

Consider about Definition \ref{definition 1}, the system is mean-square exponentially, which completes the proof.

\end{appendix}

\end{document}